\documentclass[useAMS,usenatbib]{mn2e}

\usepackage{graphicx}
\usepackage{caption}
\usepackage{times}%
\usepackage{natbib}
\usepackage{graphics}
\usepackage{epsfig}
\usepackage{array}
\usepackage{stfloats}
\usepackage{fixltx2e}
\usepackage{amsmath}
\usepackage[a4paper,colorlinks=true,pdfstartview=FitV,linkcolor=red,citecolor=blue,urlcolor=magenta]{hyperref}
\usepackage{float}

\newcommand{\gtrsim}{\ga}

\newcommand{\gsim}{\,\lower2truept\hbox{${>\atop\hbox{\raise4truept\hbox{$\sim$}}}$}\,}

\newcommand{\be}{\begin{equation}}
\newcommand{\ee}{\end{equation}}
\newcommand{\bea}{\begin{eqnarray}}
\newcommand{\eea}{\end{eqnarray}}

\renewcommand{\vec}[1]{ {\bmath #1} } 

\def\ltsima{$\; \buildrel < \over \sim \;$}
\def\simlt{\lower.5ex\hbox{\ltsima}}
\def\gtsima{$\; \buildrel > \over \sim \;$}
\def\simgt{\lower.5ex\hbox{\gtsima}}

\title[Simulating Dark Scattering]{Simulating Momentum Exchange in the Dark Sector}
\author[M.~Baldi \& F.~Simpson]{\parbox{\textwidth}{Marco Baldi$^{1,2,3}$, Fergus Simpson$^{4}$}
\\
\\$^{1}$Dipartimento di Fisica e Astronomia, Alma Mater Studiorum Universit\`a di Bologna, viale Berti Pichat, 6/2, I-40127 Bologna, Italy;
\\$^{2}$INAF - Osservatorio Astronomico di Bologna, via Ranzani 1, I-40127 Bologna, Italy;
\\$^{3}$INFN - Sezione di Bologna, viale Berti Pichat 6/2, I-40127 Bologna, Italy;
\\$^{4}$ICC, University of Barcelona (UB-IEEC), Marti i Franques 1, 08028, Barcelona, Spain.}

\def\m@th{\mathsurround=0pt }
\def\eqalign#1{\null\,\vcenter{\openup1\jot \m@th
 \ialign{\strut\hfil$\displaystyle{##}$&$\displaystyle{{}##}$\hfil
 \crcr#1\crcr}}\,}

\begin{document}
\pagerange{\pageref{firstpage}--\pageref{lastpage}} \pubyear{2011}
\maketitle
\label{firstpage}
\begin{abstract}	
\\
Low energy interactions between particles are often characterised by elastic scattering.
Just as electrons undergo Thomson scattering with photons,  dark matter particles may experience an analogous form of momentum exchange with dark energy. We investigate the influence such an interaction has on the formation of linear and nonlinear cosmic structure, by running for the first time a suite of N-body simulations with different dark energy equations of state and scattering cross sections. In models where the linear matter power spectrum is suppressed by the scattering, we find that on nonlinear scales the power spectrum is strongly enhanced. This is due to the friction term increasing the efficiency of gravitational collapse, which also leads to a scale-independent amplification of the concentration and mass functions of
halos. The opposite trend is found for models characterised by an increase of the linear matter power spectrum normalisation. More quantitatively, we find that power spectrum deviations at nonlinear scales ($k \approx 10\, h/$Mpc) are roughly ten times larger than their linear counterparts, exceeding $100\%$ for the largest value of the scattering cross section considered in the present work.
Similarly, the concentration-mass relation and the halo mass function show deviations up to $100\%$ and $20\%$, respectively, over a wide range of masses. Therefore, we conclude that nonlinear probes of structure formation might provide much tighter constraints on the scattering cross section between dark energy and dark matter as compared to the present bounds based on linear observables.
\end{abstract}

\begin{keywords}
dark energy -- dark matter --  cosmology: theory -- galaxies: formation
\end{keywords}


\section{Introduction}
\label{i}

Aside from their insensitivity to electromagnetism, the physical characteristics of dark matter and dark energy remain highly uncertain. The dark matter particle may have a mass lying anywhere in the range $10^3$ to $10^{14}$ eV, while the microphysical nature of dark energy is even less clear.  The observation that these two phenomena currently possess energy densities that are the same order of magnitude has led to speculation that they may forge a deeper connection. 
Recent measurements of the geometry of the Universe point towards an expansion history consistent with dark energy taking the form of a cosmological constant. The evolution of cosmological density perturbations provides a crucial independent test, which helps to break degeneracies between a cosmological constant and other theories of dark energy. 

Current observational data relating to density perturbations in the low redshift Universe, such as weak gravitational lensing from CFHTLenS \citep[][]{syspaper}, redshift space distortions \citep[][]{BOSSRSD2012, BlakeWigglezRSD}, and galaxy clusters \citep[][]{Xrays}, all indicate a slightly lower  amplitude of clustering than has been inferred from the Cosmic Microwave Background \citep[CMB][]{2013PlanckParams}. This tension has already been the topic of  statistical analysis \citep[][]{2014MacCrann,Beutler_etal_2014}, although the level of statistical significance hinges on assumptions such as spatial flatness. It is worth stressing that -- if confirmed -- a retardation in the growth of large scale structure is  not necessarily the work of a modified theory of gravity, but may simply arise from an interaction between the two components of the Universe which we know least about - dark matter and dark energy \cite{Simpson_Jackson_Peacock_2011}. 

Models of coupled dark energy have been extensively studied in the literature \citep[see e.g.][]{Wetterich_1995,Barrow_Clifton_2006, 2000PhRvD..62d3511A,Baldi_2011a}.   However almost all proposed models share a common feature which is that they predominantly involve energy exchange between the two fluids, with only a small degree of momentum exchange.  (The distinction between energy exchange and momentum exchange depends on the frame of reference, here we define these terms with respect to the rest frame of the dark energy fluid.)  Models with energy exchange inevitably modify both the expansion history and the growth of structure.  In this work we elaborate on a model of momentum exchange presented in \citet{simpscat}, by exploring for the first time the evolution of nonlinear structure with the use of N-body simulations. Since the energy exchange in this model is negligible, the expansion history is the same as for the non-scattering case, and is determined solely by the dark energy equation of state. 

Other forms of non-gravitational physics may also influence structure formation. Scattering between dark matter particles leads to a differential motion between baryons and dark matter. This  may become apparent via the dissipation of substructure in clusters \citep[][]{Meneghetti_etal_2001}, the formation of asymmetric tidal streams \citep[][]{Kesden_Kamionkowski_2006a}, and the sequestration of X-ray gas \citep[][]{Harvey_etal_2014}. Any interaction between dark matter and dark energy will also invoke this differential motion, and could therefore induce similar observational signatures. 

{Cosmological simulations of models characterised by some form of interaction between dark energy and dark matter have been performed by several authors \citep[see e.g.][]{Maccio_etal_2004,Baldi_etal_2010,Li_Barrow_2011,Carlesi_etal_2014a}, also including the effects of hydrodynamical forces on the uncoupled baryonic particles \citep[as e.g. in][]{Baldi_2011a,Baldi_Viel_2010,Carlesi_etal_2014b}. Several numerical studies have been devoted to the investigation of various statistical properties of the large-scale structures in the presence of such interactions, as e.g. the halo mass function \citep[][]{Baldi_Pettorino_2011,Cui_Baldi_Borgani_2012}, the clustering of galaxies in real and redshift space \citep[][]{Marulli_Baldi_Moscardini_2012,Moresco_etal_2014}, the distribution of halo satellites \citep[][]{Baldi_Lee_Maccio_2011} and of large cosmic voids \citep[][]{Li_2011,Sutter_etal_2014}, as well as weak lensing properties \citep[][]{Beynon_etal_2012,Giocoli_etal_2012,Carbone_etal_2013,Pace_etal_2014}. As mentioned above, all these investigations were carried out for interaction scenarios characterised by an energy-momentum exchange between dark energy and dark matter, while our present study is the first -- to our knowledge -- to focus on models of pure momentum exchange. }

The paper is organised as follows. In Section~\ref{sec:models} we review the modification to the dynamics of dark matter particles in the event that they experience elastic scattering with the dark energy fluid. This framework is applied to N-body simulations in Section~\ref{sec:simulations}. The results of these simulations, in terms of the matter power spectrum, halo mass function, concentration-mass relation and halo velocity dispersion are presented in Section~\ref{sec:results}. Finally in Section~\ref{sec:concl} we draw our conclusions.

\section{Dark Scattering Models}
\label{sec:models}

At sufficiently low energies the interactions between particles - elementary or composite - can invariably be described by a process of elastic scattering.  Rayleigh scattering and Thomson scattering are two prominent examples. 

Consider the case of a particle traversing an isotropic fluid whose stress energy tensor given by $T_{ab} =  \mathrm{diag}(\rho, w \rho, w \rho, w \rho)$, where $w$ is the equation of state parameter of the fluid. Provided $w\neq-1$, the particle observes a nonzero momentum flux $\bar{T}^\mu_0$, which imparts a force proportional to the scattering cross section $\sigma_c$.  It may be shown that the four-force $g^{i}$ takes the form \citep[][]{1997JApA...18...87P}:
\begin{equation}
\label{eq:four_force}
g^i = \sigma_c \left[ T^i_{\, k} u^k - u^i \left(T_{ab} u^a u^b \right) \right] \, ,
\end{equation}
in order to satisfy the condition  $g^{i} u_{i} = 0$. These terms may be expressed as
\begin{equation}
\label{eq:components}
\eqalign{
T_{ab}u^a u^b &= \rho \gamma^2 (1 + v^2 w) \, ,\cr
T_b^a u^b &= \rho \gamma(1, -vw)  \, ,
}
\end{equation}
\noindent {where $v$ is the velocity of the particle traversing the fluid and $\gamma $ is  the Lorentz factor.} This yields a force given by 
\begin{equation}
g^i  = (\gamma \mathbf{f \cdot  v} , \gamma   \mathbf{f} )   
\end{equation}
where
\begin{equation}
\label{eq:three_force}
\mathbf{f} = - (1 + w) \sigma_c  \gamma^2  \rho \mathbf{v}  \, .
\end{equation}
One cosmological example of this scenario is in the early Universe, where non-relativistic electrons experienced a retardation from the background radiation, $w_\gamma = 1/3$, given by the Thomson drag force  
\begin{equation}
\label{eq:thomson_drag}
F = -\frac{4}{3}  \sigma_T  v  \rho_{\gamma} \, ,
\end{equation}
\noindent where $\sigma_T$ is the Thomson cross section, and {$ \rho_{\gamma}$  } is the energy density of the radiation.

In this work we shall adopt the drag force given by (\ref{eq:three_force}) as the only non-gravitational force acting between the two dominant fluids in the Universe, dark matter and dark energy. 
In particular, we will consider the case of Cold Dark Matter (CDM) particles moving through the Dark Energy (DE) fluid -- such that $w = w_{\rm DE}$ -- with a DE-CDM scattering cross-section $\sigma _{c}$.

\subsection{Linear Structure Formation}

In linear perturbation theory the two fluids obey the coupled differential equations \citep[][]{simpscat}:
\begin{equation}
\label{eq:velocity_Q}
\eqalign{
\theta'_Q &= 2 H \theta_Q - a n_D \sigma_c \Delta \theta + k^2 \phi + k^2 \frac{\delta_Q}{1+w}   \, , \cr
\theta'_c  &= - H \theta_c + \frac{\rho_Q}{\rho_c} (1+w) a n_D \sigma_c \Delta \theta + k^2 \phi  \, ,
}
\end{equation}
\noindent where $n_D$ is the proper number density of dark matter particles, $\Delta \theta \equiv \theta_Q - \theta_c$ is the velocity contrast,  ($\theta _{i}$ being the divergence of the velocity perturbations for the field $i$), and  where we have assumed the dark energy sound speed $c_s^2 = 1$. The prime denotes a derivative with respect to time. The other perturbation equations remain in their canonical form, i.e.:
\begin{equation}
\label{eq:eq3}
\eqalign{
\delta'_Q =& - \left[ \left(1+w\right) + 9\frac{H^2}{k^2} \left(1-w^2\right) \right] \theta_Q   \cr + & ~ 3 (1+w) \phi' - 3H(1-w) \delta_Q   ,
}
\end{equation}
\begin{equation}
\label{eq:eq4}
\delta'_c = - \theta_c + 3 \dot{\phi}  .
\end{equation}

The gravitational potential $\phi$ is sourced by the Poisson equation. Baryons may be included as a third fluid with no coupling term. This slightly weakens the modification to the growth rate in cosmologies with coupled fluids \cite{Simpson_Jackson_Peacock_2011}, but otherwise leaves the functional form unchanged, so we shall neglect their influence in this work.
 
\subsection{Nonlinear Structure Formation}

In the context of our N-body simulations, our equation of motion for the dark matter fluid element is modified by the additional term involving the CDM particle mass $m_{\rm CDM}$, its velocity $v$, the scattering cross section $\sigma _{c}$, the dark energy equation of state $w_{\rm DE}$ and the dark energy density $\rho _{\rm DE}$, according to the equation:
\begin{equation}
\label{new_drag}
\dot{v} = - (1 + w_{\rm DE}) \sigma_c \frac{ v c \, \rho_{\mathrm{DE}} }{m_{\mathrm{CDM}}}
\end{equation}
where $c$ is the speed of light.

As we are assuming a DE sound speed $c_{s}^{2} = 1$, we expect DE perturbations to be  dumped within the cosmic horizon so that the DE density and velocity field is approximately homogeneous. Therefore, we shall neglect the influence of dark energy perturbations within the simulation as we will concentrate on scales falling well within the horizon. We have numerically verified that this assumption has little impact on the derived growth rate even for larger scales.
\begin{table}
\begin{center}
\begin{tabular}{cc}
\hline
Parameter & Value\\
\hline
$H_{0}$ & 67.8 km s$^{-1}$ Mpc$^{-1}$\\
$\Omega _{\rm M} $ & 0.308 \\
$\Omega _{\rm DE} $ & 0.692 \\
$ \Omega _{b} $ &0.0482 \\
\hline
${\cal A}_{s}$ & $2.215 \times 10^{-9}$\\
$n_{s}$ & 0.966\\
\hline
\end{tabular}
\end{center}
\caption{The set of cosmological parameters adopted in the present work, consistent with the latest results of the Planck collaboration \citep[][]{Planck_016}. { Here $n_{s}$ is the spectral index of primordial density perturbations while ${\cal A}_{s}$ is the amplitude of primordial density perturbations.}}
\label{tab:parameters}
\end{table}

\subsection{Energy Exchange}

The net rate of change in the dark energy density may be expressed as a sum of the contributions from the dark matter and the adiabatic expansion. 
\begin{equation}
\frac{d  \rho_Q}{d a} = \rho_Q \left[1+w(a)\right]   \left[ \sigma_c n_0 \frac{v^2}{a^4H} - \frac{3}{a} \right] \, ,
\end{equation}
where $n_0$ indicates the number density of dark matter particles at the present day. Consider the evolution of $\rho_Q$ during the era of radiation domination. The velocity dispersion $ v^2 \simeq {\rm const.}$ and $H \propto a^{-2}$, thus for a constant equation of state $w$ we find:
\begin{equation}
\rho_Q = \rho_{Q0} a^{-3(1+w)} e^{-\frac{1+w}{a} \frac{ \sigma_c n_0 v_{eq}^2}{H_{eq}}}
\end{equation}

This suggests that the dark energy grows extremely quickly in the early universe, when the dark matter particles had a very high number density. The energy density reaches a maximum when either (a) the adiabatic decay takes over; or (b) the energy transfer becomes large enough to saturate the kinetic energy of dark matter; or (c) the equation of state transitions to $w \simeq -1$.  The nature of this energy transfer will be investigated in greater detail in future work. Here we shall focus on the consequences of momentum exchange during matter-domination, where the available kinetic energy density is low and thus we can safely assume any change in the dark energy density is negligible.

\section{The Simulations}
\label{sec:simulations}

In order to explore the nonlinear effects of the Dark Scattering scenario described above, 
we have run a series of CDM-only cosmological N-body simulations by means of a suitably modified version
of the widely-used TreePM code {\small GADGET-3} \citep[][]{gadget}. Our modified version implements the additional drag term described
by Eq.~\ref{new_drag} for CDM simulation particles. For a generic system of $N$ particles in an expanding background the full acceleration equation experienced by the $i$-th particle then becomes:
\begin{equation}
\label{acceleration}
\dot{\vec{v}}_{i} = -\left[ 1 + A \right] H \vec{v}_{i} + \sum_{j \ne i}\frac{Gm_{j}\vec{r}_{ij}}{|\vec{r}_{ij}|^{3}}
\end{equation}
where $\vec{r}_{ij}$ is the distance between the $i$-th and the $j$-th particle. The extra scattering term $A$ is defined as:
\begin{equation}
\label{drag_1}
A\equiv \left( 1 + w_{\rm DE}\right) \sigma_{c} \frac{c}{m_{\rm CDM}} \frac{3 \Omega _{\rm DE}}{8\pi G} H
\end{equation}
and adds to the standard cosmological friction term.

This extra drag force depends on three free quantities: the DE equation of state parameter $w_{\rm DE}$, the DE-CDM scattering cross section $\sigma_c$, and the CDM particle mass $m_{\rm CDM}$. While the former could be in general an arbitrary function of redshift, $w(z)$, the latter two quantities are  more naturally modelled as dimensional constants. The overall magnitude of the drag force depends only on their ratio, so we can define the combined quantity 
\begin{equation}
\label{xi}
\xi \equiv \frac{c\cdot \sigma_{c}}{m_{\rm CDM}}
\end{equation}
with dimensions of $[{\rm bn} \cdot {\rm c}^{3}/{\rm GeV}]$, as the characteristic parameter of our models, such that Eq.~\ref{drag_1} becomes:

\begin{equation}
\label{drag_2}
A\equiv \left( 1 + w\right) \frac{3 \Omega _{\rm DE}}{8\pi G} H \xi
\end{equation}
In the present work, we restrict our investigation to the simplified (and rather unrealistic) case of a constant DE equation of state, $w_{\rm DE}={\rm const.}$ {This is done in order to reduce the number of free parameters in this first numerical investigation so to focus on the effects of the DE-CDM scattering using a relatively low number of simulations.} More realistic equations of state for the DE component will be discussed in a forthcoming paper, for which the present simplified analysis will serve as a guideline to select reasonable combinations of possible $w_{\rm DE}(z)$ and $\xi $. 

\begin{table}
\begin{center}
\begin{tabular}{cccc}
\hline
Run & $w_{\rm DE}$  &  
$\xi [{\rm bn} \cdot {\rm c}^{3}/{\rm GeV}]$ &
$\sigma_{8}(z=0)$\\
\hline
$\Lambda $CDM & $-1$ & -- & 0.826 \\
w09-xi0 & $-0.9$ & $0$ & 0.803\\
w09-xi10 & $-0.9$ & $10$ & 0.792\\
w09-xi30 & $-0.9$ & $30$ & 0.770\\
w09-xi50 & $-0.9$ & $50$ & 0.750\\
w11-xi0 & $-1.1$ & $0$ & 0.845\\
w11-xi10 & $-1.1$ & $10$ & 0.852\\
w11-xi30 & $-1.1$ & $30$ & 0.865\\
w11-xi50 & $-1.1$ & $50$ & 0.879\\
\hline
\end{tabular}
\end{center}
\caption{The suite of cosmological N-body simulations considered in the present work, with their main physical and numerical parameters.}
\label{tab:models}
\end{table}

Under these assumptions, we consider three possible values of the DE equation of state parameter  $w_{\rm DE} = \left\{ -1.1\,, -1\,, -0.9\right\}$, and four possible values of the parameter $\xi = \left\{ 0\,, 10\,, 30\,, 50\right\} \, {\rm bn} \cdot {\rm c}^{3}/{\rm GeV}$. As clearly shown by Eq.~\ref{drag_2}, for the case of $w_{\rm DE}=-1$ (corresponding to a cosmological constant and therefore having the same expansion history as the standard $\Lambda $CDM cosmology) the additional drag force vanishes, irrespective of the value of $\xi $, and the model fully recovers the standard $\Lambda $CDM scenario. We are therefore left with 9 distinct models, summarised in Table~\ref{tab:models}, for each of which we have performed a cosmological N-body simulation within a periodic box of 250 comoving Mpc$/h$ a side, filled with $512^{3}$ particles with mass $m_{c}=9.95 \times 10^{9}$ M$_{\odot}/h$. The gravitational softening has been set to $\epsilon _{g} = 12$ kpc$/h$, corresponding to roughly 1/40$th$ of the mean inter-particle separation. 

\begin{figure}
\includegraphics[width=\columnwidth]{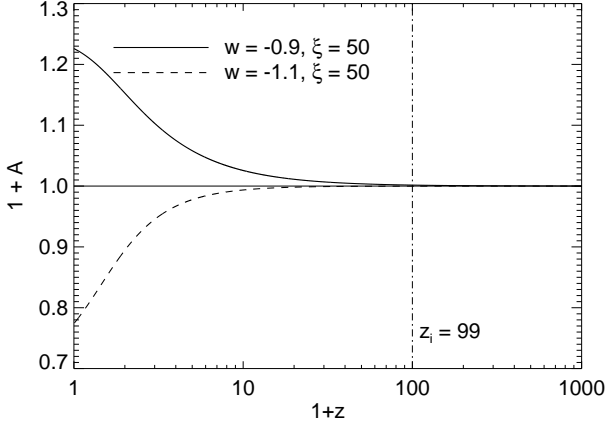}
\caption{{ The full friction coefficient $(1+A)$ as a function of redshift for the two most extreme scenarios under consideration in the present work, namely the two cases of $\xi = 50\, {\rm bn} \cdot {\rm c}^{3}/{\rm GeV}$ for $w_{\rm DE} = -0.9$ ({\em solid}) and $w_{\rm DE} = -1.1$ ({\em dashed}). As one can see in the plot, for these models the effect of the DE-CDM scattering on the overall friction term in the perturbations equations is negligible (below $1\%$) for $z\gtrsim 50$, which makes our numerical setup fully consistent. For more realistic scenarios such as EDE models we expect that this condition will no longer hold.}}
\label{fig:drag_force}
\end{figure}

All simulations started from the same initial conditions at $z_{i}=99$. The initial conditions have been generated by randomly displacing particles from a homogeneous {\em ``glass"} distribution \citep[][]{Davis_etal_1985} according to the Zel'dovich approximation \citep[][]{Zeldovich_1970} for a $\Lambda $CDM cosmology with parameters based on the latest results of the Planck satellite mission \citep[][]{Planck_016}, which are summarised in Table~\ref{tab:parameters}. In doing so, we are discarding any possible effect of the DE-CDM scattering at redshifts higher than $z_{i}$. This approximation appears to be fully justified for the models under investigation, since the effective  drag force is proportional to the combination $\Omega _{\rm DE}H$ which rapidly vanishes at high redshifts for our set of cosmologies. This is clearly shown in Fig.~\ref{fig:drag_force} where we display the magnitude of the overall friction term $(1+A)$ as a function of redshift for the two most extreme models considered in the present paper. The effect of the scattering is found to be negligible at $z > z_{i}$. Clearly, for more realistic scenarios as e.g. Early Dark Energy models \citep[see e.g.][]{EDE,EDE2}, for which the DE fractional density $\Omega _{\rm DE}$ does not vanish at high redshifts, this approximation will have to be dropped. 

\begin{figure}
\includegraphics[width=\columnwidth]{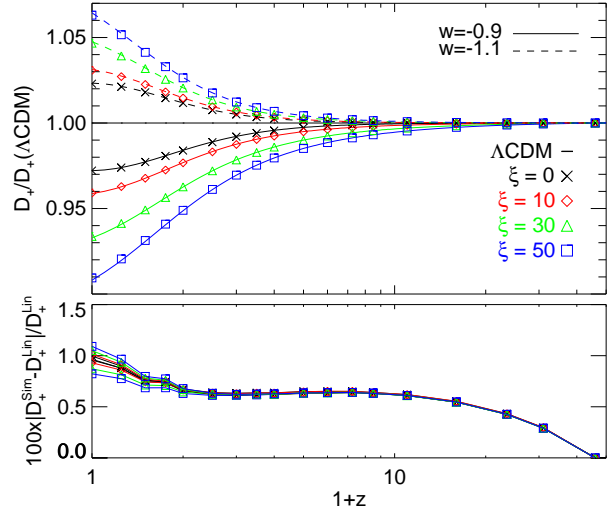}
\caption{{ {\em Top}: The ratio of the linear growth factor to the $\Lambda $CDM case for the different models under investigation extracted from our suite of simulations (open symbols) and computed by integrating the linear equations ~\ref{eq:velocity_Q},\ref{eq:eq4} (solid and dashed lines). {\em Bottom}: The accuracy of our N-body code in capturing the linear growth of the different models: the relative difference between the linear and the simulated growth functions never exceeds $1.2\%$.}}
\label{fig:growth_factor}
\end{figure}

Although starting from the same initial conditions, the various models will show a different evolution due to both the different cosmological expansion history and the different impact of the DE-CDM scattering. Therefore, the amplitude of linear perturbations at $z=0$ will be different, resulting in a different value of $\sigma _{8}$ as shown in the last column of Table~\ref{tab:models}. We have tested that our modified code correctly captures this different evolution of linear perturbations by comparing the linear growth factor extracted from the simulations with the theoretical expectation obtained by numerically integrating Eqs.~\ref{eq:velocity_Q},\ref{eq:eq4}. The comparison is shown in Fig.~\ref{fig:growth_factor}, where we show in the upper panel the ratio of the growth factor $D_{+}(z)$ to the standard $\Lambda $CDM case both for the theoretical expectations in the $w_{\rm DE} = -0.9$ and $w_{\rm DE} = -1.1$ models (solid and dashed lines, respectively) and for the simulations results (open symbols). As on can see from the plot, the two relative deviations agree extremely well from $z=45$ (corresponding to the first snapshot of our simulations) to $z=0$. In the lower panel we display the absolute value of the percent relative difference between the simulated and the expected linear growth. The plot shows that the accuracy of the code is at the $\sim 1\%$ level for all the models and redshifts.

\section{Results}
\label{sec:results}

We now move to discuss the main results of our set of simulations, consisting of a series of basic large-scale structure statistics that will be systematically compared both with the standard $\Lambda $CDM cosmology and with the constant-$w_{\rm DE}$ models with no DE-CDM scattering.

\subsection{Large-scale matter distribution}

\begin{figure}
\begin{center}
\includegraphics[scale=0.25]{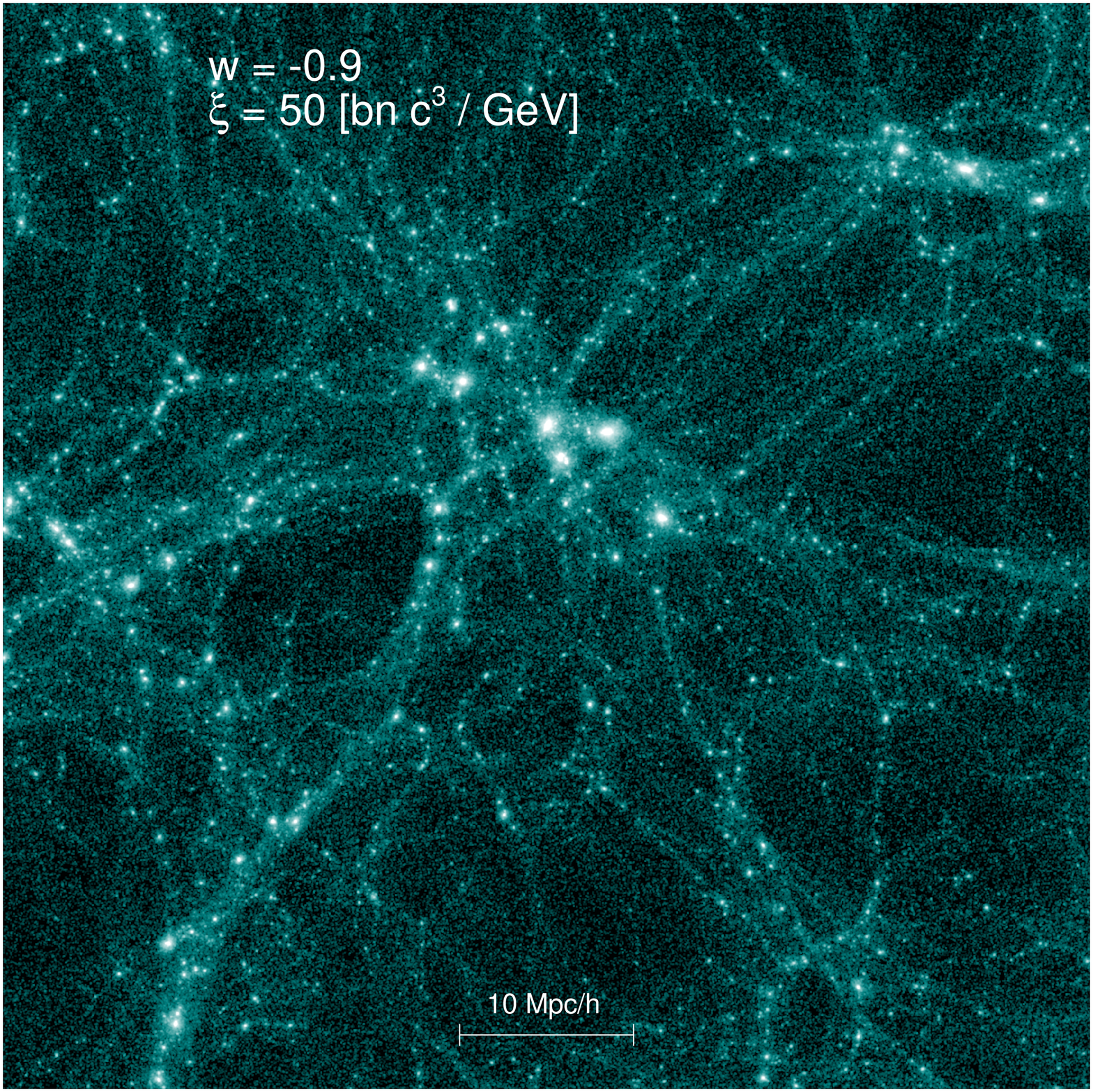}
\includegraphics[scale=0.25]{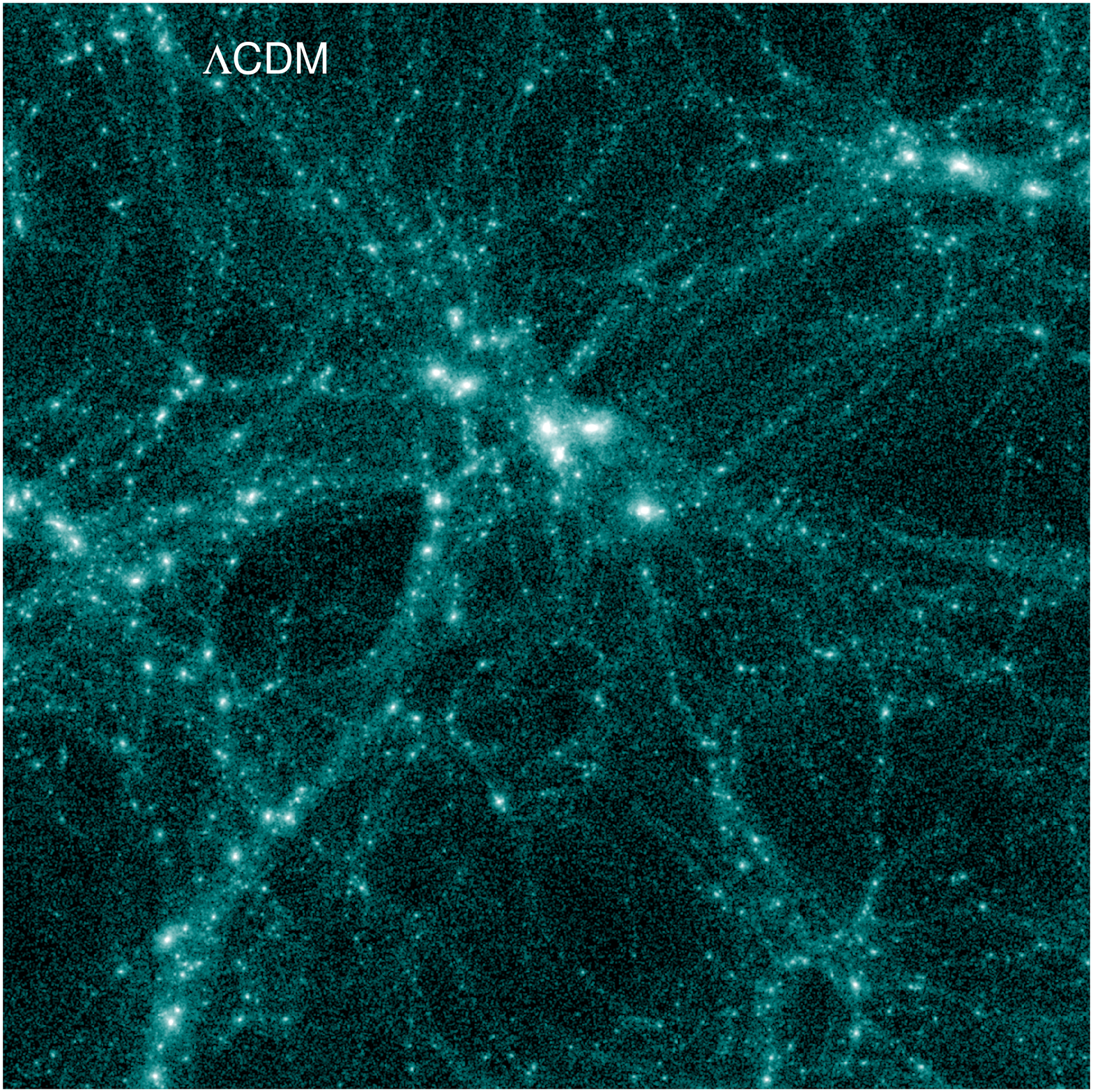}
\includegraphics[scale=0.25]{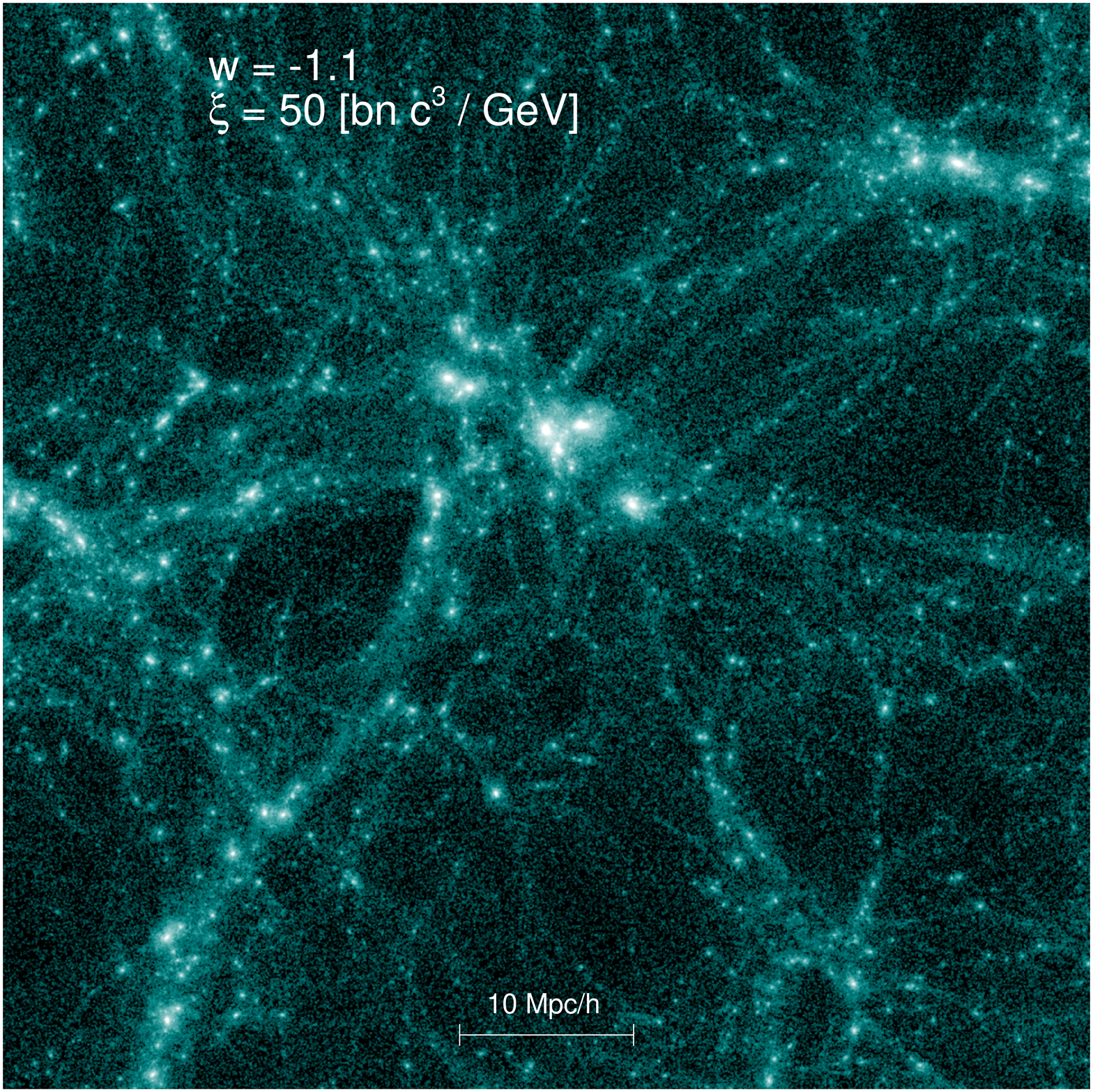}
\end{center}
\caption{{ Density slices in the reference $\Lambda $CDM cosmology ({\em middle}) and in the two $\xi = 50 [{\rm bn} \cdot {\rm c}^{3}/{\rm GeV}]$ models with $w=-0.9$ ({\em top}) and $w=-1.1$ ({\em bottom})}}
\label{fig:CDM}
\end{figure}

We begin with the visual inspection of the CDM distribution in a slice of thickness $37.5$ Mpc$/h$ through the simulation box. In Fig.~\ref{fig:CDM} we show the projected CDM density field at $z=0$ in the standard $\Lambda $CDM cosmology ({\em middle panel}) and in the two most extreme dark scattering scenarios, namely the $\xi = 50\, [{\rm bn} \cdot {\rm c}^{3}/{\rm GeV}]$ models with $w_{\rm DE} = -0.9$ and $w_{\rm DE} = -1.1$ ({\em upper} and {\em lower} panels, respectively). The size of the region shown in the figures is $52.5$ Mpc$/h$ a side (i.e. one fourth of the whole simulation box). 

By looking at the three images, one can notice how the global shape of the large-scale cosmic web is the same in the three models, reflecting their common initial conditions. At the same time, from this visual inspection it is already possible to notice that the three models are characterised by a different level of evolution of the large-scale structures, with the $w_{\rm DE} = -1.1$ clearly showing a stronger clustering of the most prominent density peaks, while the $w_{\rm DE} = -0.9$ cosmology has a lower clustering than the $\Lambda $CDM case. This is consistent with the general picture of  the DE-CDM scattering enhancing or suppressing the linear growth of structures in the two cases.

\subsection{The nonlinear matter power spectrum}

\begin{figure*}
\includegraphics[scale=0.3]{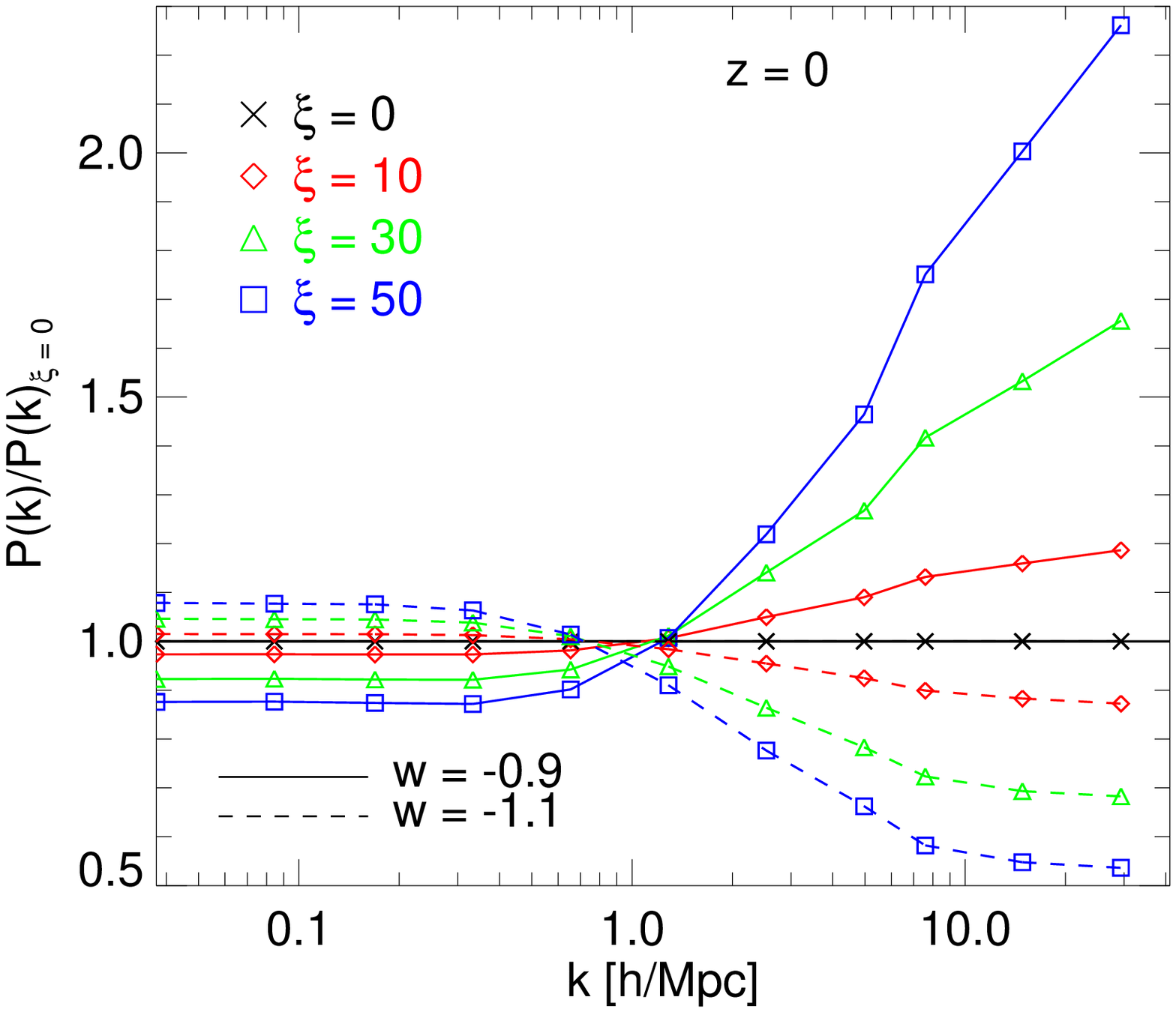}
\includegraphics[scale=0.3]{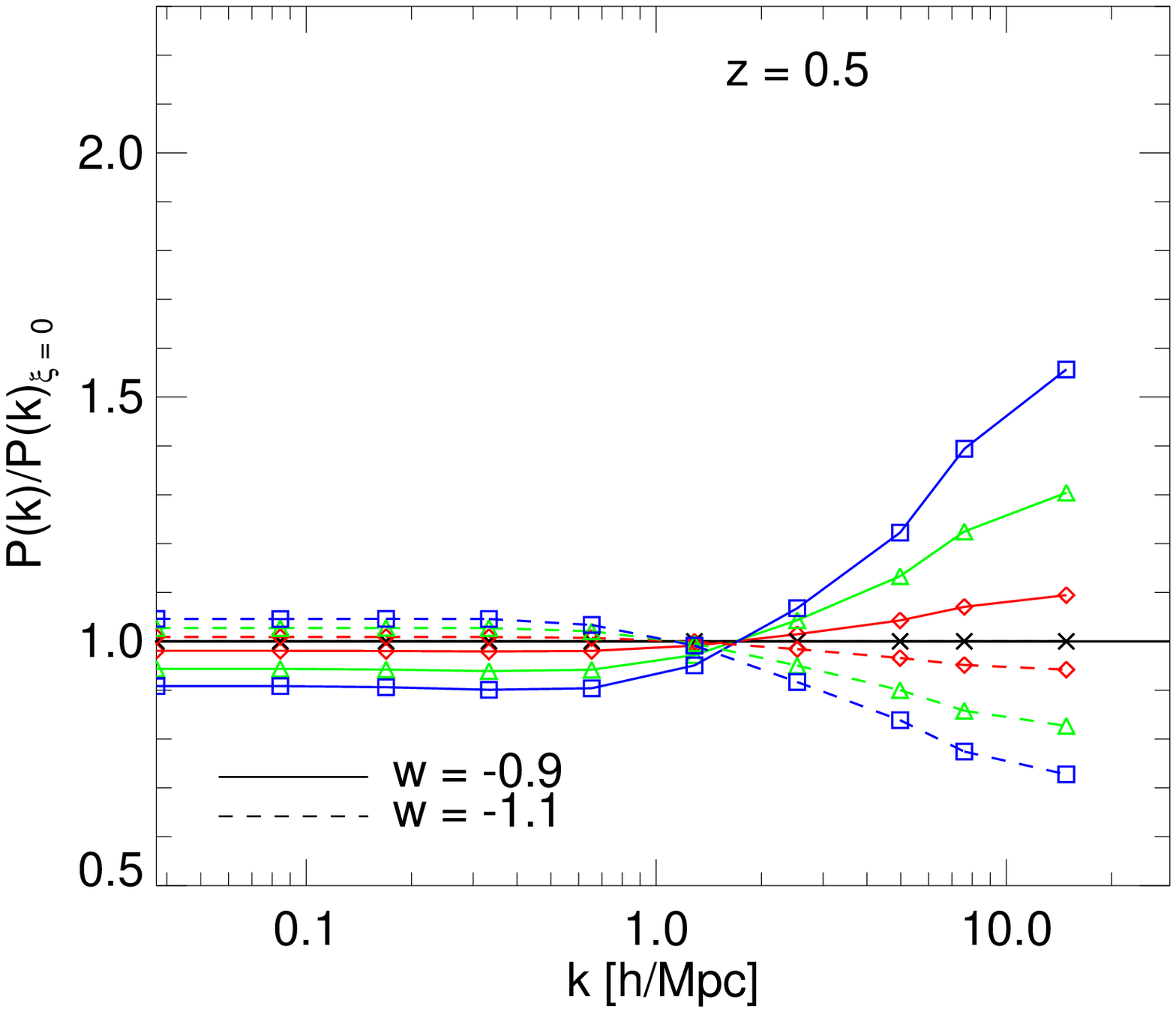}
\includegraphics[scale=0.3]{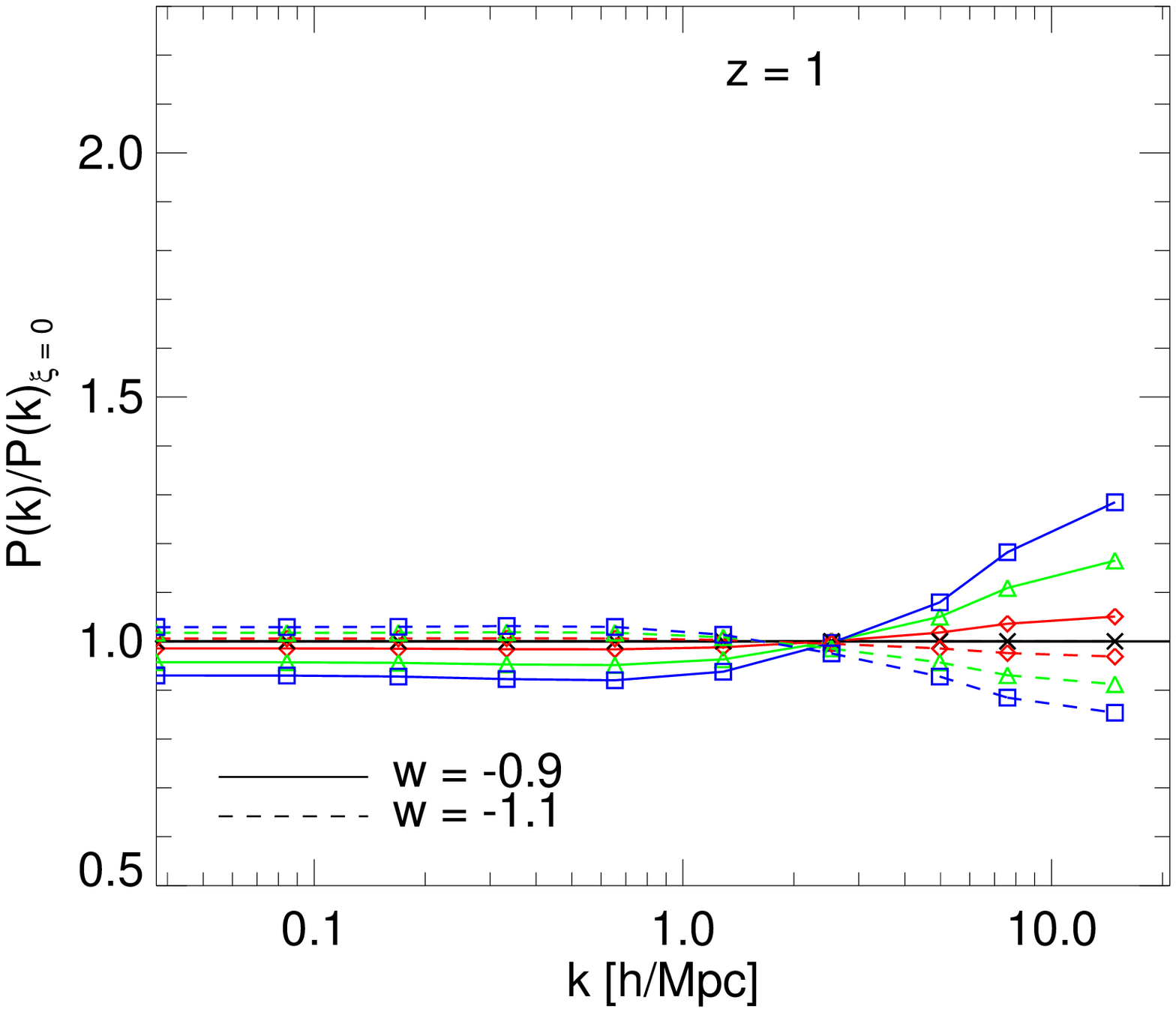}
\caption{{ The nonlinear matter power spectrum ratio to the $\xi = 0$ case for models with both $w_{\rm DE} = -0.9$ ({\em solid lines}) and $w_{\rm DE} = -1.1$ ({\em dashed lines}) at three different redshifts $z=0$ ({\em left}), $z=0.5$ ({\em middle}), and $z=1$ ({\em right}). The different colours and open symbols refer to the different values of the parameter $\xi $. As the figures show, the DE-CDM scattering results in two opposite effects at linear and nonlinear scales: while at the largest scales we observe a weak and scale-independent suppression (enhancement) of the spectrum normalisation for $w_{\rm DE} = -0.9$ ($w_{\rm DE} = -1.1$), corresponding to the different expected values of $\sigma _{8}$ for the different models, at smaller scales we find a strong scale-dependent enhancement (suppression) of power for the same classes of models, respectively. The transition scale between the two behaviours lies in the range $0.7-3\, h/$Mpc depending on the model and on the redshift.}}
\label{fig:power_ratio}
\end{figure*}

For all the simulations of our suite we extract the nonlinear matter power spectrum through a Cloud-in-Cell mass assignment to a cubic cartesian grid with the same spacing of the PM mesh used for the large-scale integration with {\small GADGET}, i.e. $512^{3}$ grid nodes over the whole simulation box. This provides a measurement of the power spectrum up to the Nyquist frequency of the grid $k_{\rm Ny} = \pi N/L = 6.43$ Mpc$/h$. Beyond this frequency we estimate the power spectrum by means of the folding method of \citet{Jenkins_etal_1998,Colombi_etal_2009} and we smoothly interpolate the two estimations around $k_{\rm Ny}$. The total power spectrum obtained in this way is then truncated at the frequency where the shot noise reaches 20\% of the measured power.

\begin{figure*}
\includegraphics[scale=0.3]{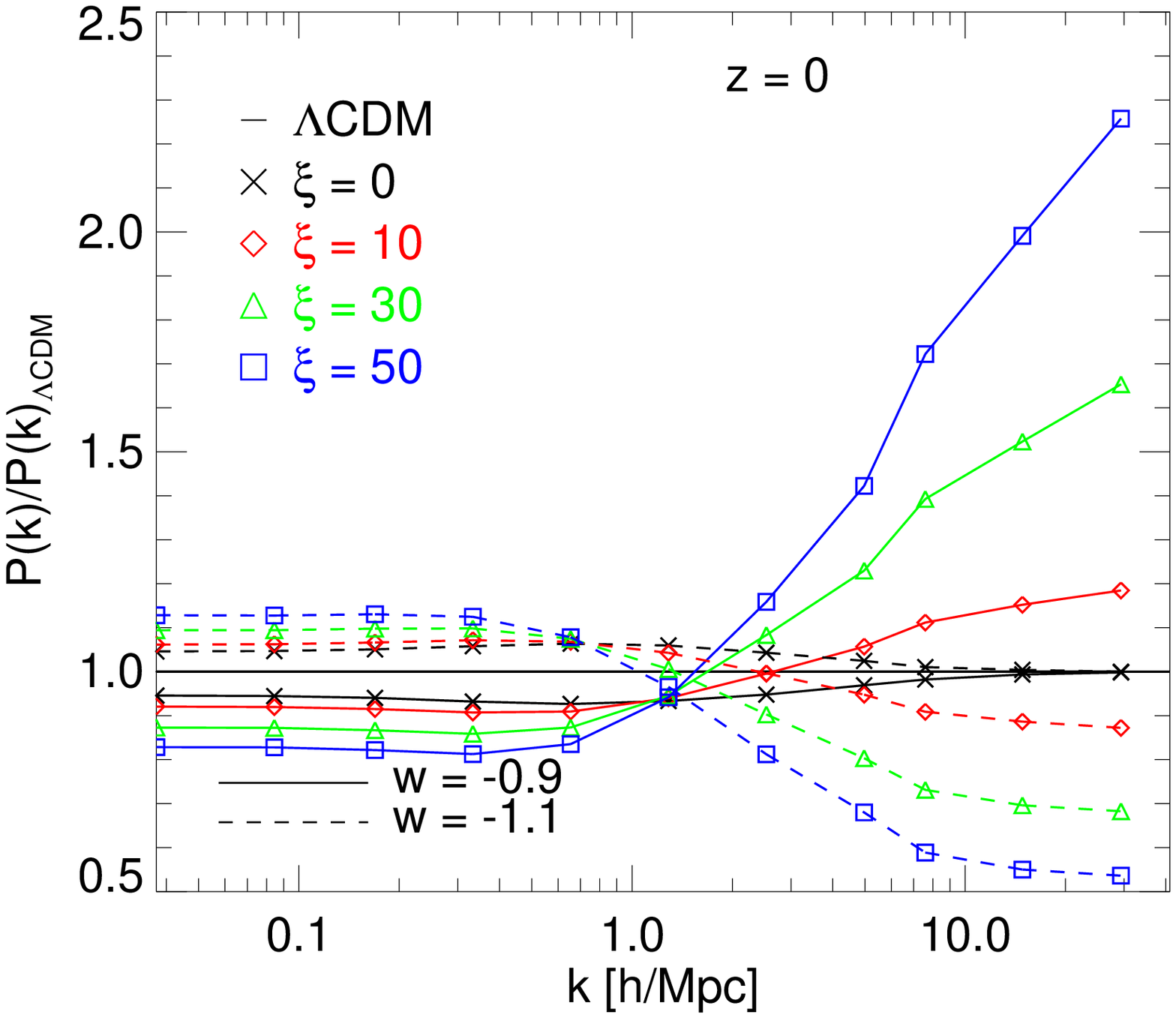}
\includegraphics[scale=0.3]{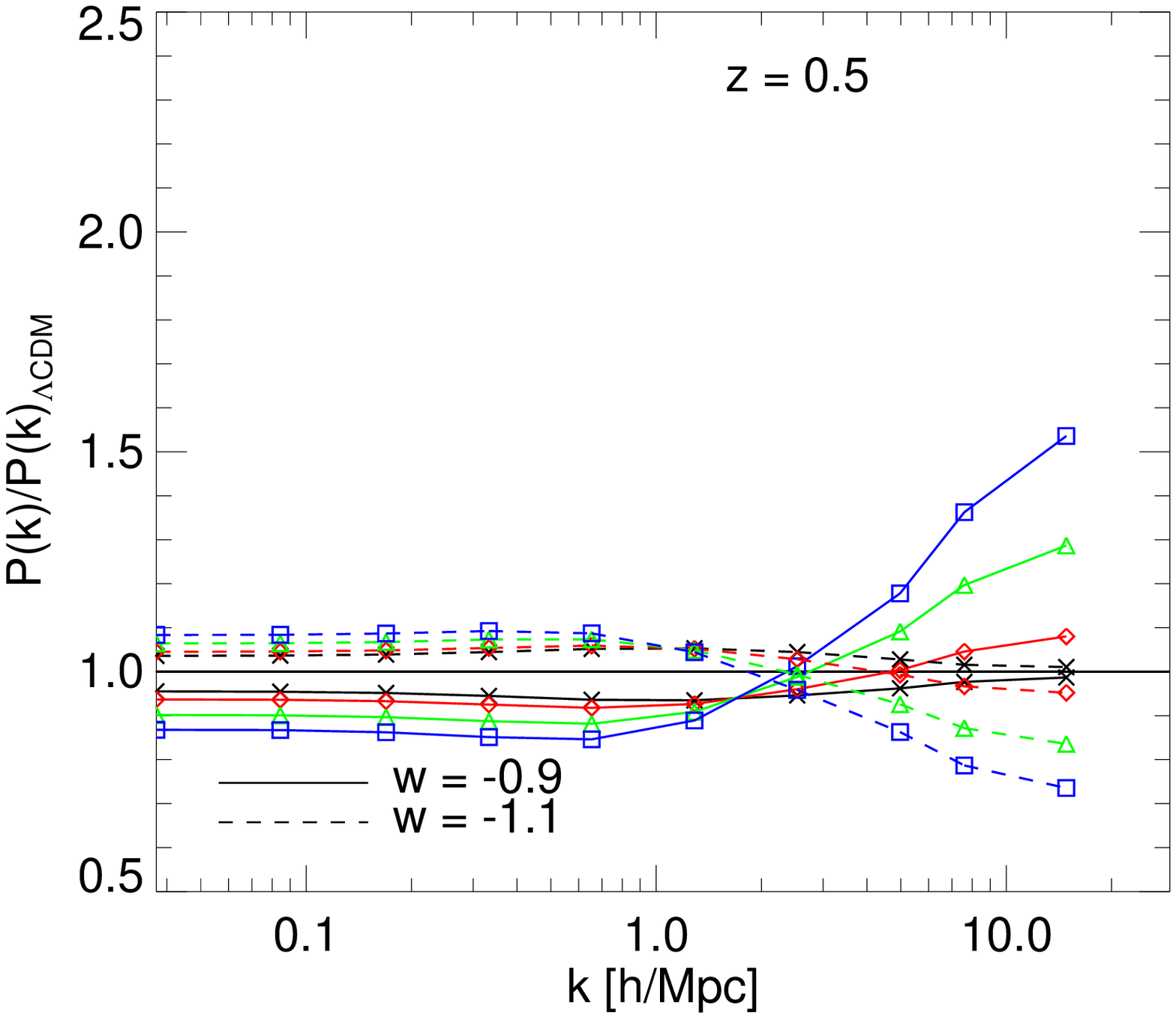}
\includegraphics[scale=0.3]{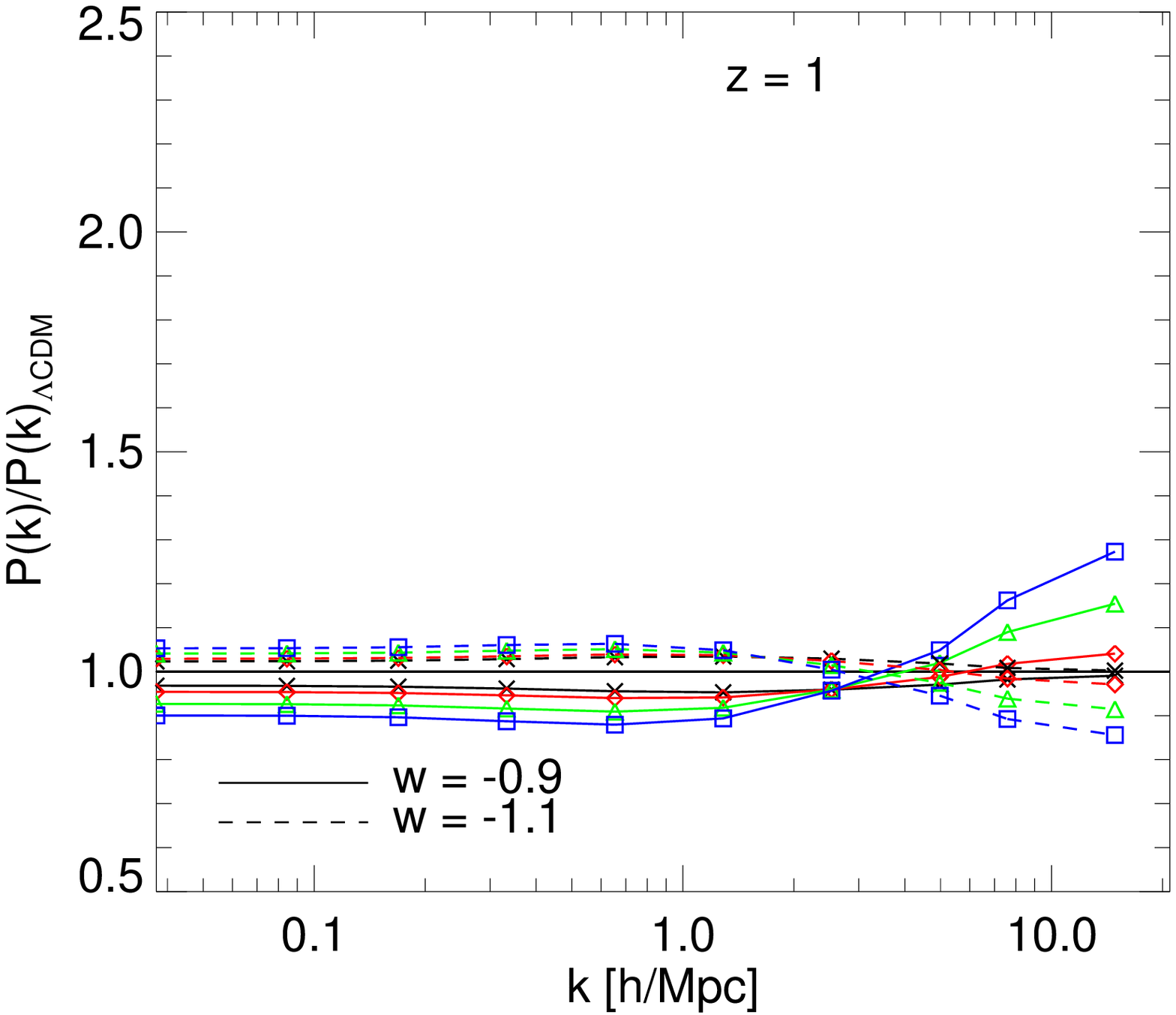}
\caption{{ The nonlinear matter power spectrum ratio to the reference $\Lambda$CDM model at three different redshifts $z=0$ ({\em left}), $z=0.5$ ({\em middle}), and $z=1$ ({\em right}). The line styles, colours, and symbols are the same as in Fig.~\ref{fig:power_ratio}. Also in this case it is possible to observe a transition between the linear and the nonlinear impact of the DE-CDM scattering on the measured power.}}
\label{fig:power_ratio_LCDM}
\end{figure*}

In Fig.~\ref{fig:power_ratio} we display the ratio of the full nonlinear matter power spectra of all the simulations with $w_{\rm DE}\neq -1$ to the model with identical expansion history and no scattering between DE and CDM (i.e. to the $\xi = 0$ case), and for three different redshifts $z=0$ ({\em left}), $z=0.5$ ({\em middle}), and $z=1$ ({\em right}). As the figure shows, on linear scales the effect of the DE-CDM scattering is to suppress (enhance) the power for $w_{\rm DE} > -1$ ($w_{\rm DE} < -1$) cosmologies, with a maximum deviation from the non-scattering model of $\approx 12\%$ at $z=0$, $\approx 10\%$  at $z=0.5$, and $\approx 8\%$ at $z=1$.  This confirms the expectations obtained through linear theory, as one should expect also by looking at the test shown in Fig.~\ref{fig:growth_factor}. 

However, at nonlinear scales -- which are evaluated here for the first time -- the situation appears significantly different. First of all, the sign of the deviation is reversed with respect to linear scales, showing an enhancement of the nonlinear power for $w_{\rm DE} > -1$ and a suppression for $w_{\rm DE} < -1$. The transition between the two regimes lies in the range $k\sim 0.7 - 3\, h/$Mpc for the various models and redshifts. Furthermore, the amplitude of the effect is much larger in the nonlinear regime than observed for the linear case, with a maximum deviation exceeding $100\%$ for the most extreme scenarios at the smallest scales probed by our present resolution. Also, the deviation from the non-scattering case shows a strong scale-dependence in the nonlinear regime, suggesting that the amplitude of the effect might keep increasing at even smaller scales. This result clearly shows how the nonlinear regime might provide much tighter constraints on the scattering cross section $\sigma _{c}$ between DE and CDM (or more precisely on the ratio $\xi $ between the cross section and the CDM particle mass) as compared to what obtained so far using only linear probes \citep[][]{simpscat}.

Similarly to what displayed in Fig.~\ref{fig:power_ratio}, we show in Fig.~\ref{fig:power_ratio_LCDM} the ratio of the full nonlinear matter power spectrum of the various cosmologies to the $\Lambda $CDM case, thereby including in the comparison both the effect of the DE-CDM scattering and of the specific expansion history associated with the two different values of the equation of state parameter $w_{\rm DE}$. As one can see from the figure, the overall effect is qualitatively similar to what observed in the comparison with the $\xi = 0$ case for a fixed expansion history. This shows that the impact of the background cosmic evolution is subdominant with respect to the DE-CDM scattering, in particular at highly nonlinear scales, even for rather extreme choices on the DE equation of state as the ones adopted in this study. Therefore, the nonlinear effects of the DE-CDM scattering appear as a distinctive and prominent feature of the model, independently from our ignorance about the nature of the DE component and its dynamical evolution. Remarkably, while the amplitude of the deviation at the smallest scales seems to be roughly insensitive to the background expansion history assumed as a reference, the linear deviation appears larger when comparing to the expected $\Lambda $CDM result, showing how in the linear regime the effects of a  $w_{\rm DE}={\rm const.}$ expansion history and of the associated DE-CDM scattering modify the growth of perturbations in the same direction. This would further increase the significance of this signature when comparing to real observational data.

\begin{figure*}
\includegraphics[scale=0.3]{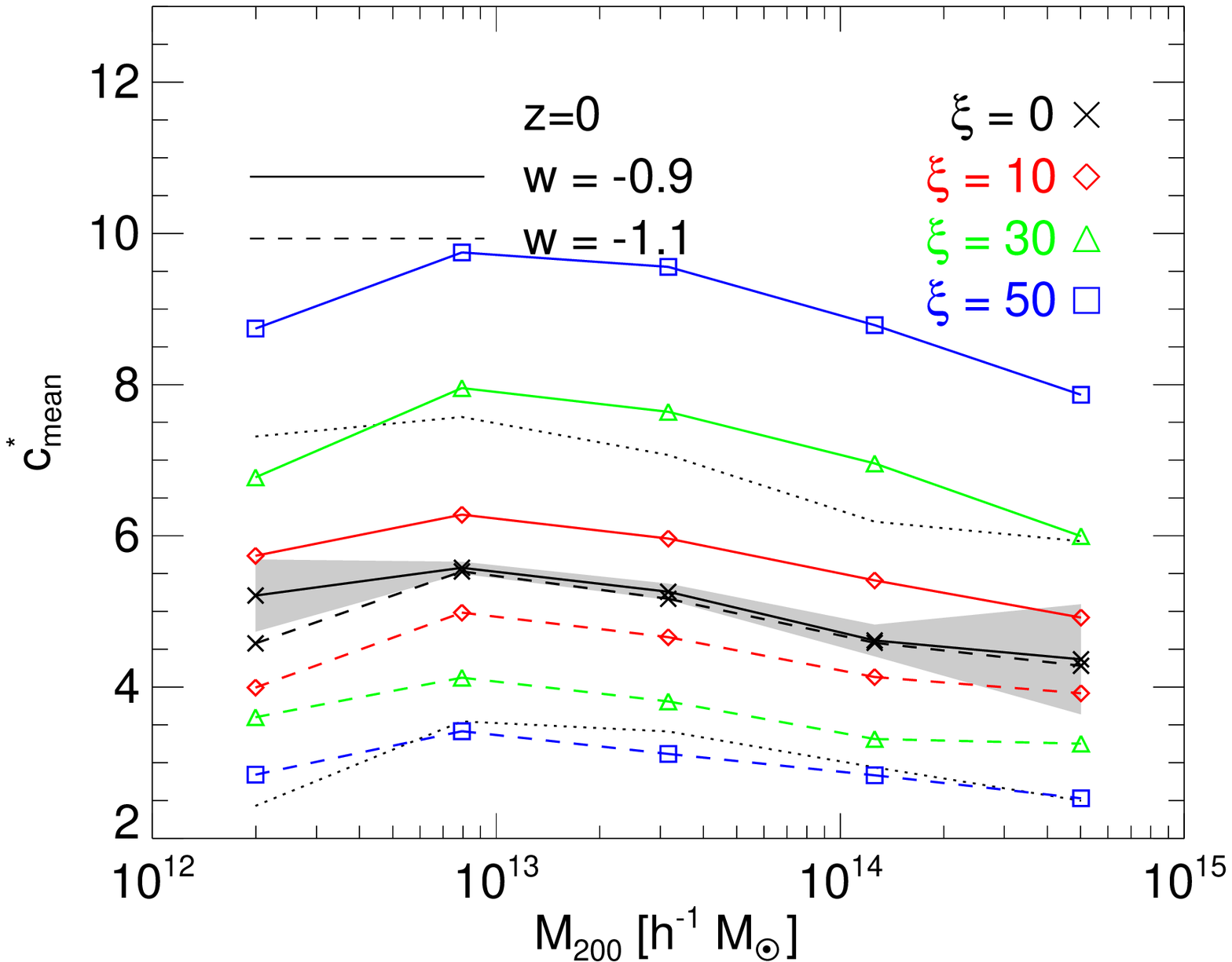}
\includegraphics[scale=0.3]{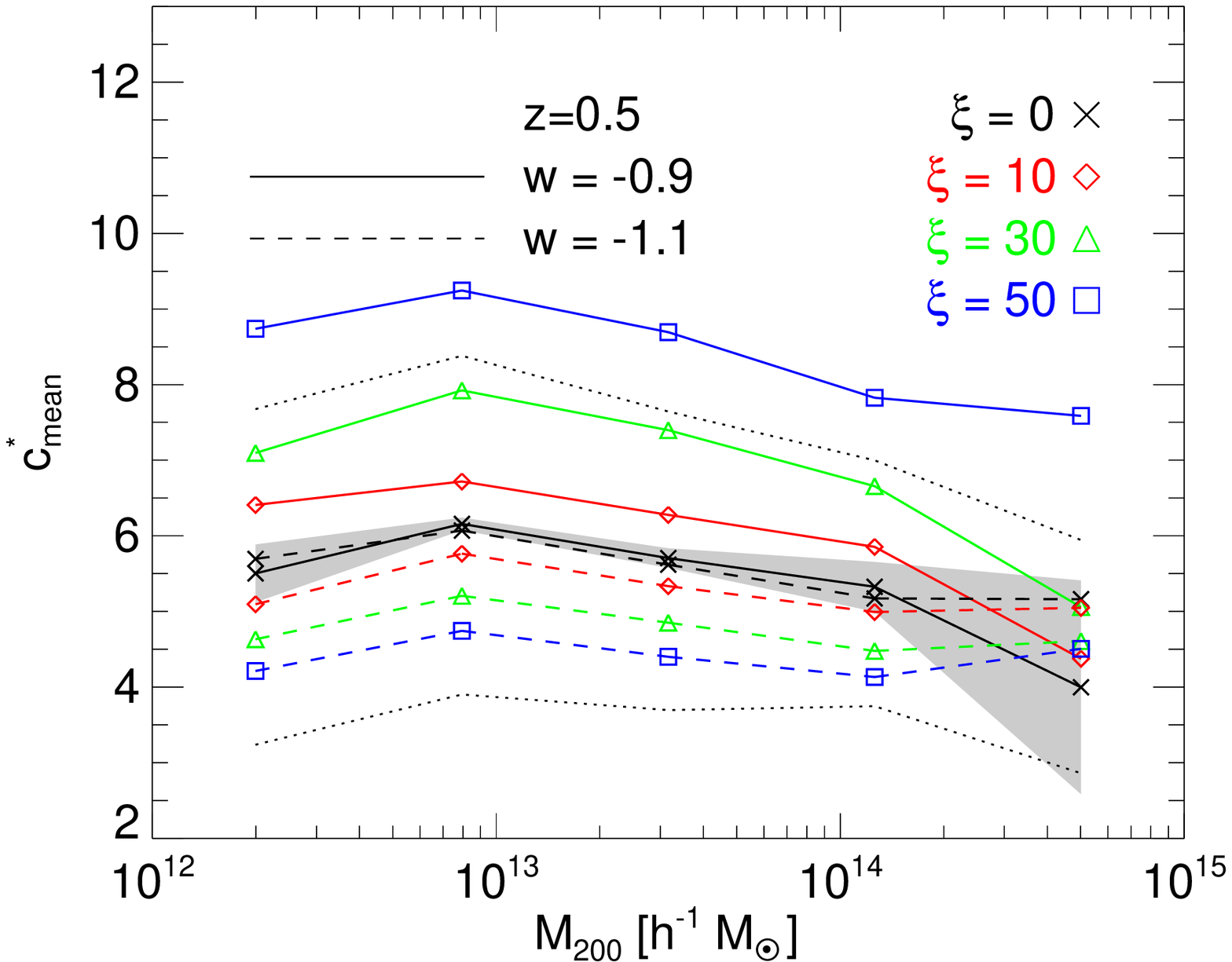}
\includegraphics[scale=0.3]{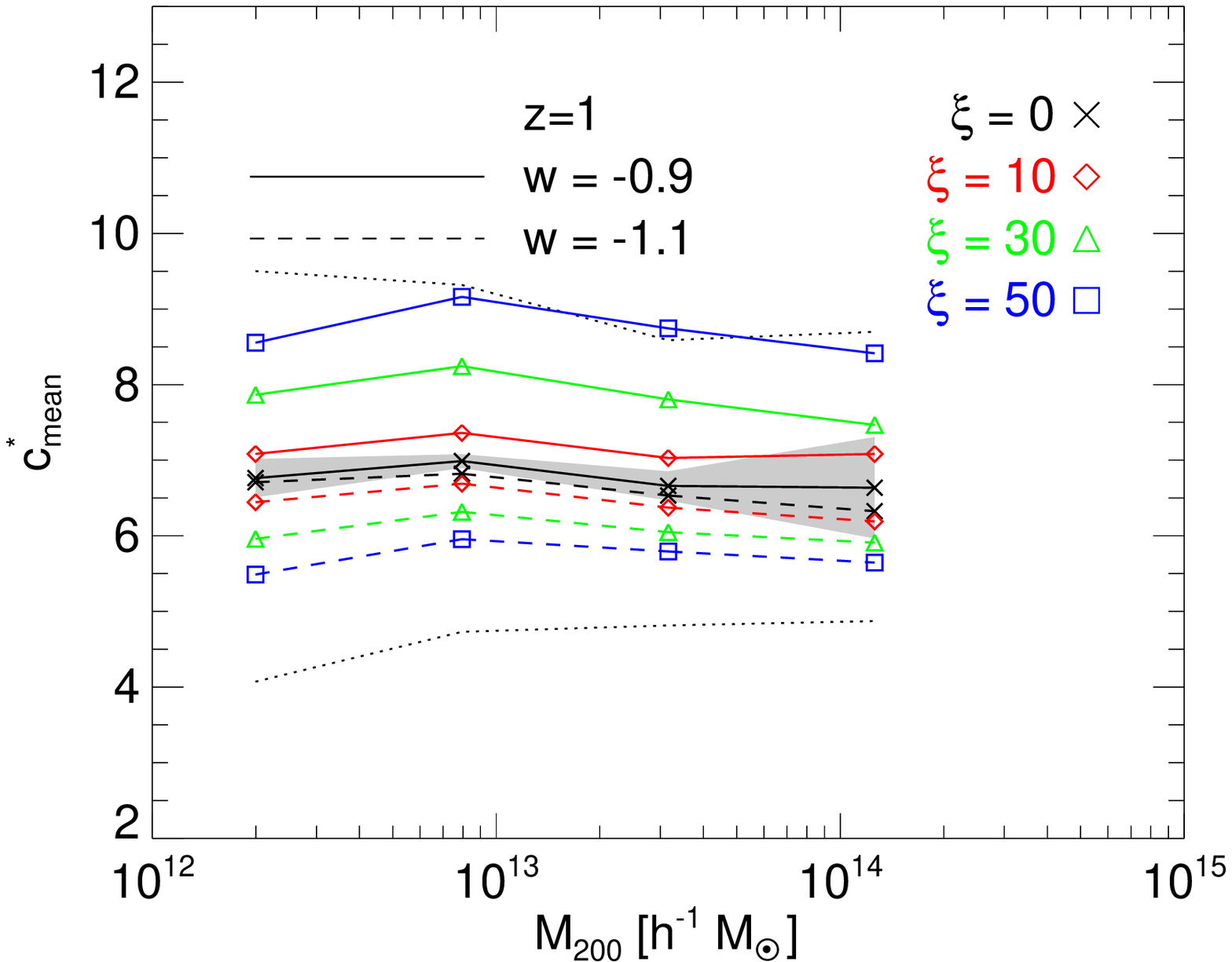}
\caption{{ The $c^{*}-M_{200}$ relation for all the models with $w_{\rm DE} \neq -1$ considered in the present work, at three different redshifts $z=0$ ({\em left}), $z=0.5$ ({\em middle}), and $z=1$ ({\em right}). Solid lines refer to the $w_{\rm DE}=-0.9$ cosmologies while dashed lines refer to $w_{\rm DE}=-1.1$ models, while the different colours and open symbols refer to the different values of the parameter $\xi $. As the plots show, $w_{\rm DE} = -0.9$ gives rise to higher concentrations for increasing values of $\xi $, while lower concentrations are obtained for $w_{\rm DE} = -1.1$.}}
\label{fig:concentrations}
\end{figure*}

\begin{figure*}
\includegraphics[scale=0.3]{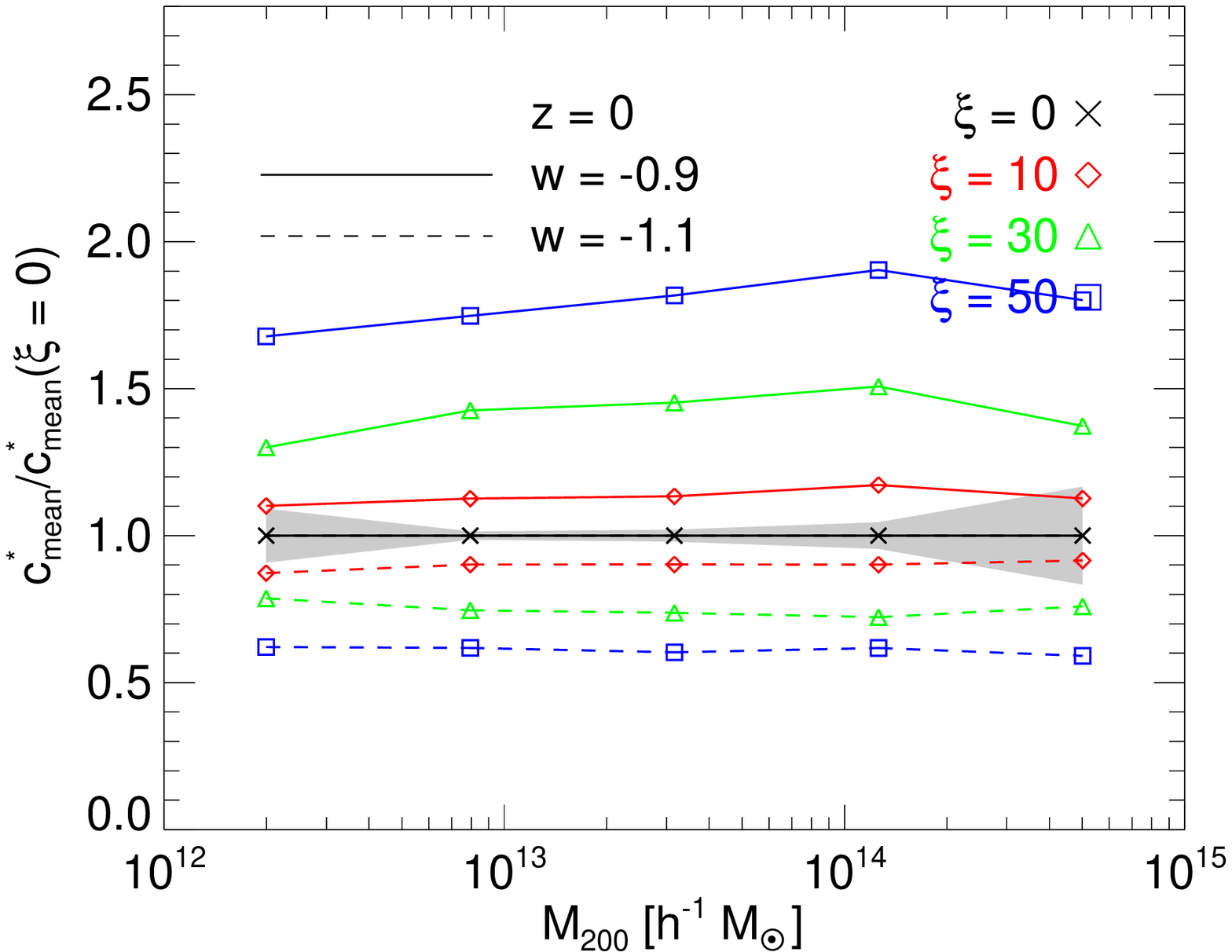}
\includegraphics[scale=0.3]{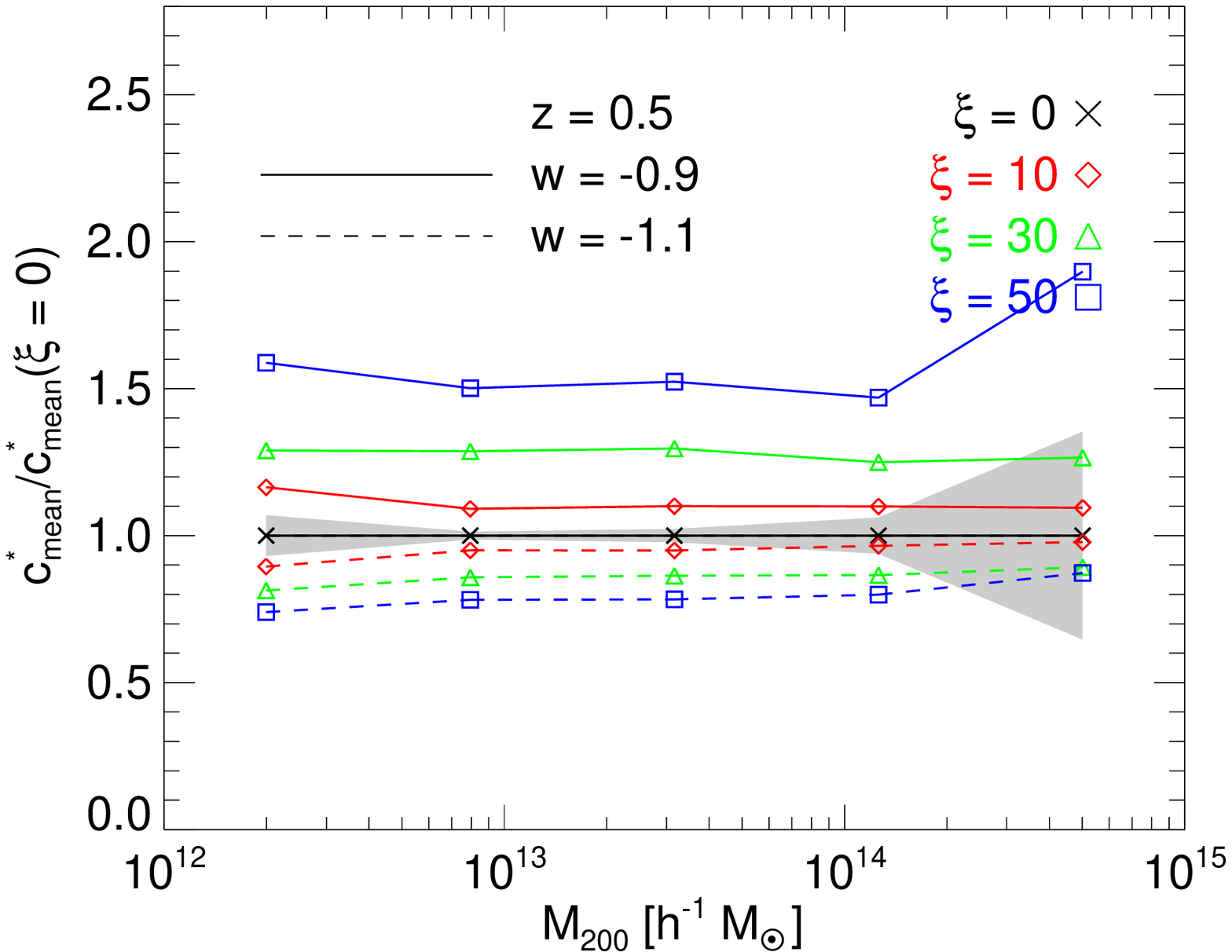}
\includegraphics[scale=0.3]{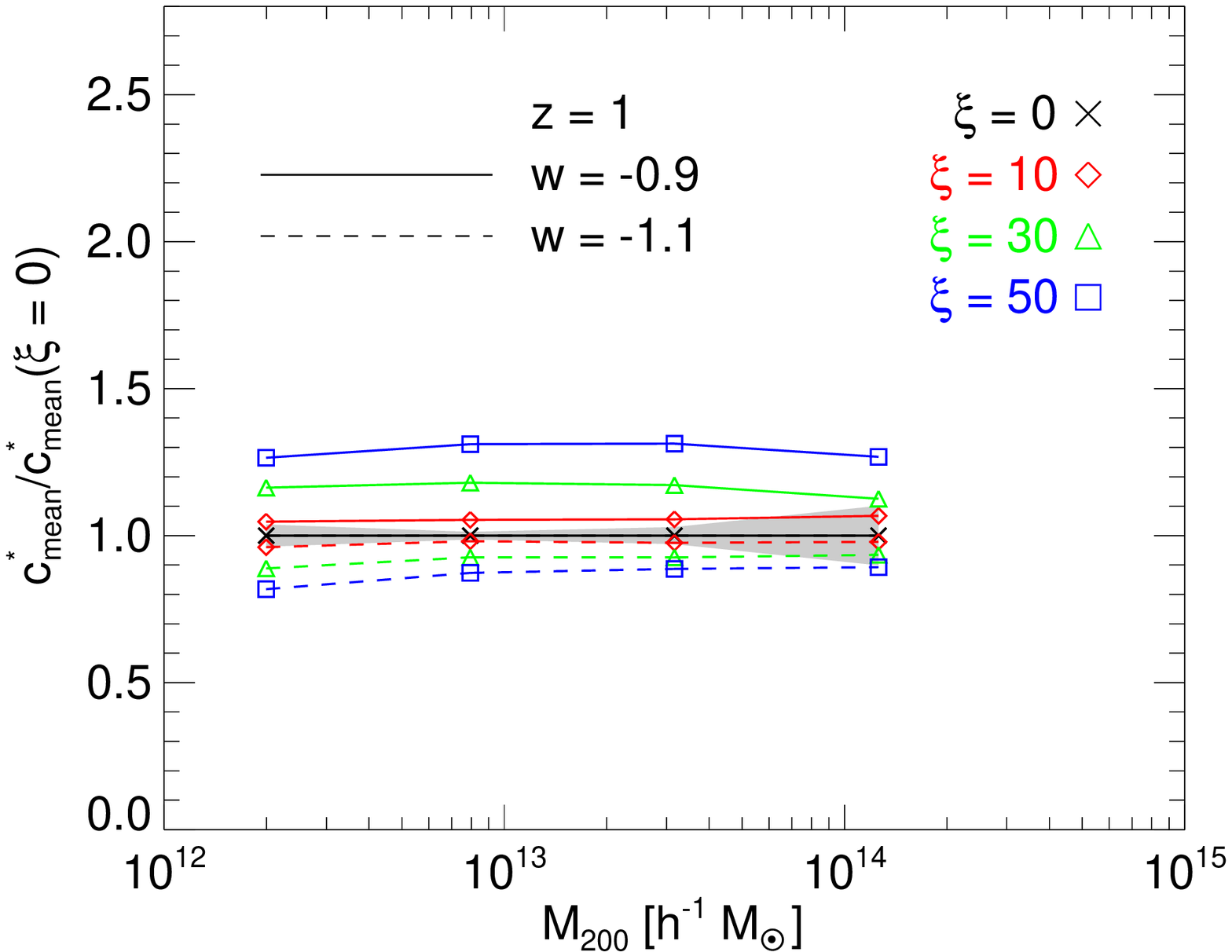}
\caption{{ The ratio of the $c^{*}-M_{200}$ relation to the $\xi = 0$ case for both values of the equation of state at three different redshifts $z=0$ ({\em left}), $z=0.5$ ({\em middle}), and $z=1$ ({\em right}). Line styles, colours,  and symbols are the same as in Fig.~\ref{fig:concentrations}}}
\label{fig:concentration_ratio}
\end{figure*}

These modification to the power spectrum can be readily interpreted as the effect of the energy-momentum dissipation due to the DE-CDM scattering. This is expected to alter the collapse and the virialisation process of gravitationally bound structures at small scales. More specifically, while in the linear regime the velocity field is always aligned with the spatial gradient of the gravitational potential, such that the drag term associated with the DE-CDM scattering always acts in the same direction of the gravitational acceleration, this is no longer true in the nonlinear regime (i.e. after shell crossing) when the collapsing structures start to gain angular momentum. Therefore, while in the linear regime an extra friction (drag) will necessarily suppress (enhance) the growth of structures, in the nonlinear regime the main effect would be to lower (increase) the kinetic energy of bound structures thereby altering their virial equilibrium and causing them to contract, in the case of a friction, or to expand in the case of a drag. 

This explains the opposite trend of the nonlinear matter power spectrum in the different models and the transition scale between suppression and enhancement coinciding with the transition between linear and nonlinear scales. Interestingly, an analogous effect has already been discussed in the context of interacting DE cosmologies by \citet{Baldi_etal_2010,Baldi_2011b} where a similar type of drag term appears in the perturbations equations as a consequence of momentum conservation. However, while in interacting DE cosmologies such an extra drag is only one amongst several other modifications of the dynamical equations, and its impact is therefore mitigated by other competing effects, in the dark scattering models under investigation in the present work it represents the only effect at play, thereby maximally displaying its impact in the large-scale matter distribution. As we will see below, the nonlinear behaviour of the extra drag characterising dark scattering cosmologies will have a significant impact also on the structural properties of CDM halos such as their concentration and their velocity dispersion.
This result illustrates how the nonlinear regime of structure formation might be used to place much tighter constraints on the ratio $\xi $ between the DE-CDM scattering cross section $\sigma _{c}$ and the CDM particle mass $M_{\rm CDM}$ as compared with the bounds that can be derived using only linear observables.

\subsection{Halo concentrations}

For all the simulations of our suite we have identified particle groups by means of a Friends-of-Friends (FoF) algorithm with linking length $\ell = 0.2\times \bar{d}$, where $\bar{d}$ is the mean inter-particle separation.
Furthermore, for each FoF group we have identified gravitationally bound substructures by means of the {\small SUBFIND} algorithm \citep[][]{Springel_etal_2001} and we associated to the main substructure of each FoF halo a spherical overdensity mass $M_{200}$  defined as the mass enclosed in a sphere of radius $R_{200}$ centred on the particle with the minimum gravitational potential such that the mean density within $R_{200}$ corresponds to $200$ times the critical density of the universe, $\rho_{crit} \equiv 3H^{2}/8\pi G$. 

For each halo in our sample we then compute the halo concentrations $c^{*}$ following the method devised in \citet{Aquarius} as:
\begin{equation}
\frac{200}{3}\frac{c^{*3}}{\ln (1+c^{*}) - c^{*}/(1+c^{*})} = 7.213~\delta _{V}
\end{equation}
with $\delta _{V}$ defined as:
\begin{equation}
\delta _{V} = 2\left( \frac{V_{max}}{H_{0}r_{max}}\right) ^{2}
\end{equation}
where $V_{max}$ and $r_{max}$ are the maximum rotational velocity of the halo and the radius at which this velocity peak is located, respectively. 

In Fig.~\ref{fig:concentrations} we show the average concentrations obtained with this method within five logarithmically equispaced mass bins as a function of the bin mass $M_{200}$ for all the models under investigation, and at the usual three different redshifts. The grey shaded area in each panel indicates the statistical Poissonian error based on the abundance of halos in each bin of the reference simulation. 

As the plot clearly shows, the DE-CDM scattering determines a significant enhancement (suppression) of the normalisation of the $c^{*}-M_{200}$ relation for $w_{\rm DE} > -1$ ($w_{\rm DE}<-1$). The effect is maximum at low redshifts and for the largest value of $\xi $, reaching a factor of $\approx 2$ in both directions for $\xi = 50$ [bn $\cdot c^{3}/$GeV]. This is more evident by looking at Fig.~\ref{fig:concentration_ratio}, where we display the ratio of the average concentration $c^{*}_{mean}$ in each mass bin to the reference case of no DE-CDM scattering ($c^{*}_{mean}(\xi =0)$) with fixed background expansion history. Also in this case, as for the comparison of the halo mass function, the deviation from the reference case does not show any clear dependence on the halo  mass. This is therefore a further peculiar feature of dark scattering models, which also predict a significant impact on the average structural properties of CDM halos besides the above mentioned effects on the large-scale matter distribution.

Since the drag force experienced by the dark matter particles scales with their velocity, particles within larger halos experience a greater force. However over a given period of time, the fractional reduction in velocity $\Delta v / v $ is independent of $v$, and therefore each halo is subject to the same proportional change in concentration.

These results are consistent with the effects observed on the nonlinear matter power spectrum and with the physical interpretation provided above for the transition between the linear and the nonlinear behavior of an extra-drag term in the dynamical evolution of CDM particles. More specifically, our findings on the halo concentrations, and in particular their evolution with redshift, clearly show how the virial equilibrium of collapsed halos is continuously altered at low redshifts by the dissipation (for $w_{\rm DE} > -1$) or the injection (for $w_{\rm DE} < -1$) of kinetic energy from (or into) the systems, resulting in or expansion of the halos, respectively, with matter moving towards (or out of) the halo core.

\subsection{The halo mass function}

\begin{figure*}
\includegraphics[scale=0.3]{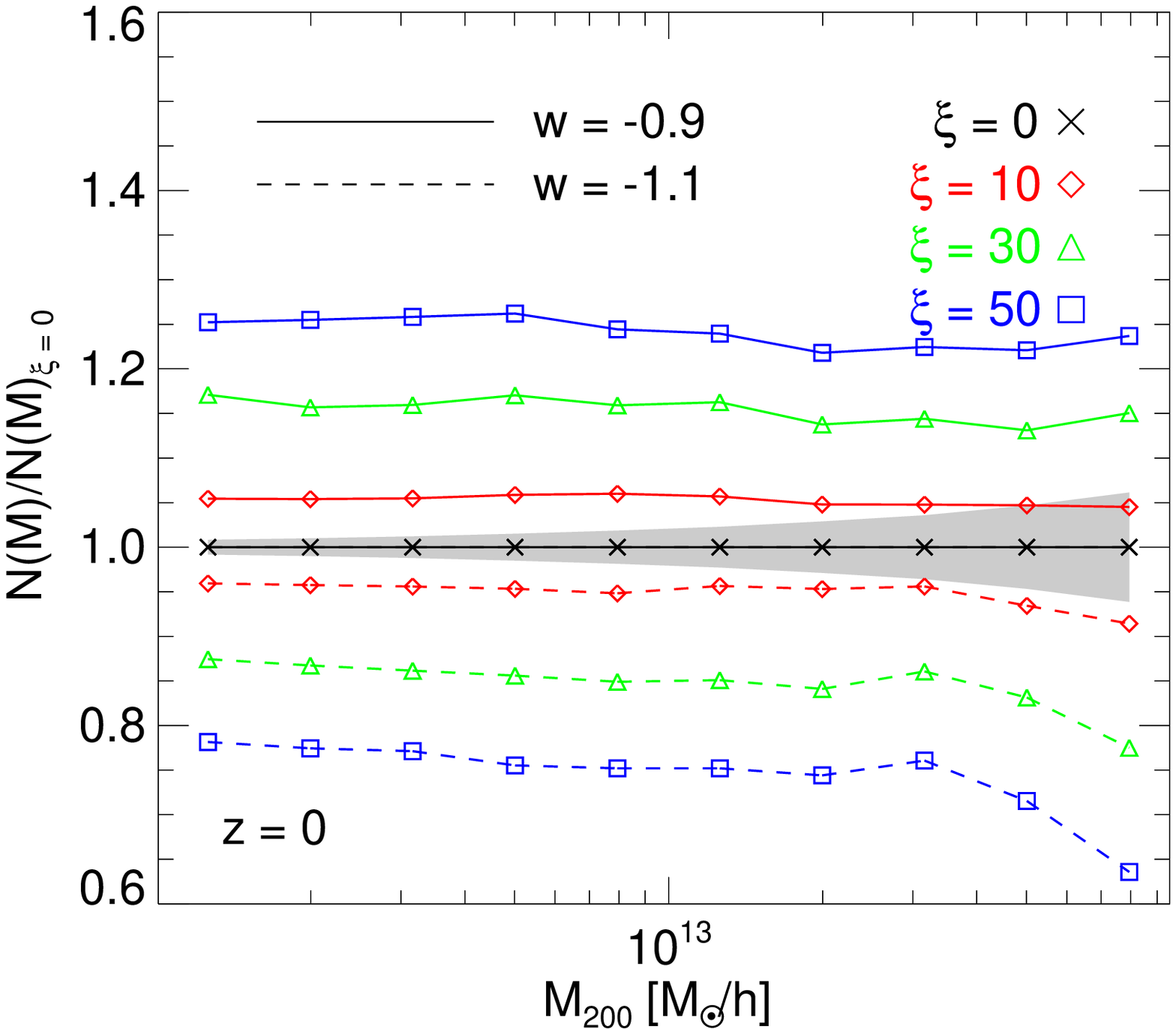}
\includegraphics[scale=0.3]{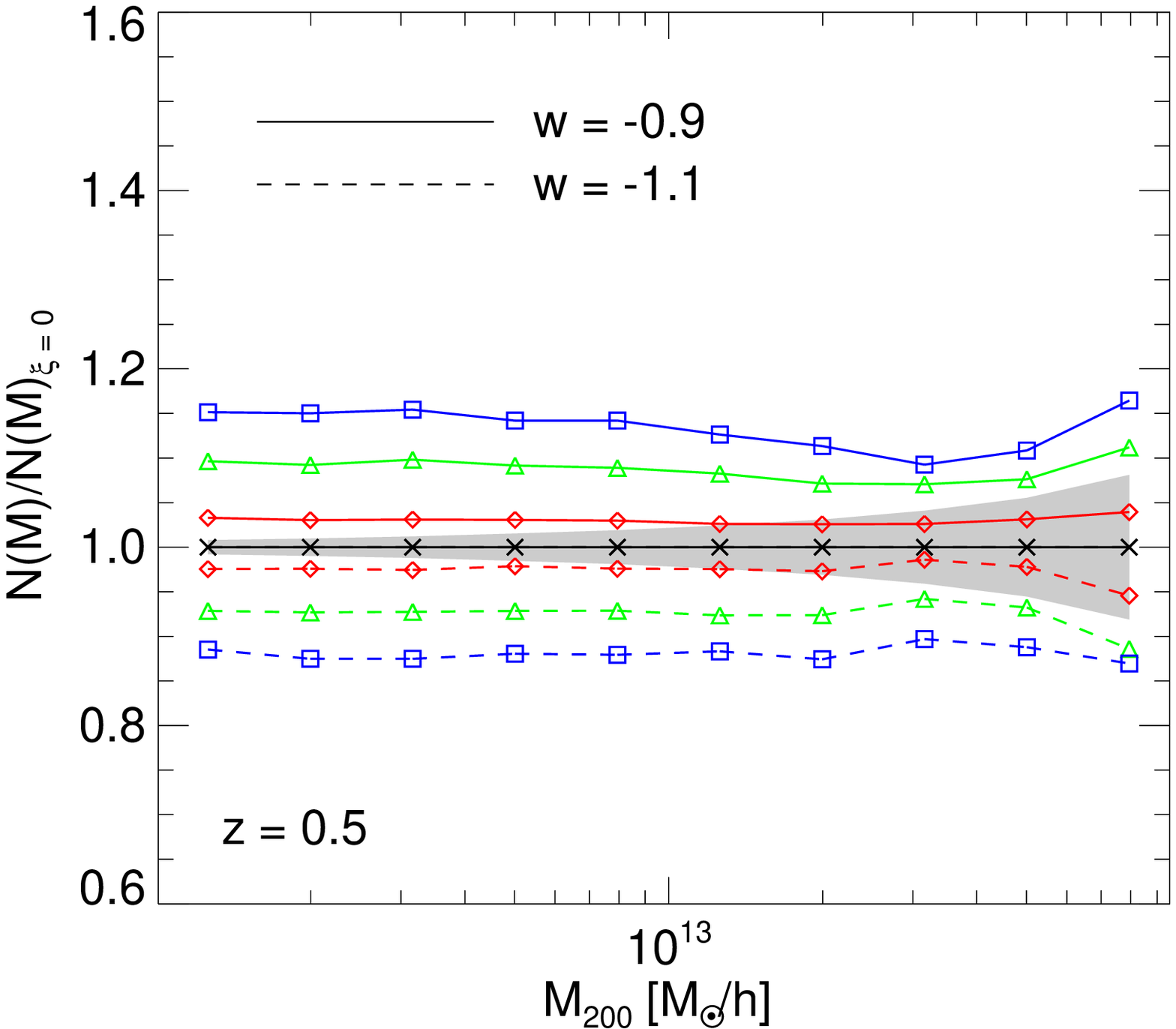}
\includegraphics[scale=0.3]{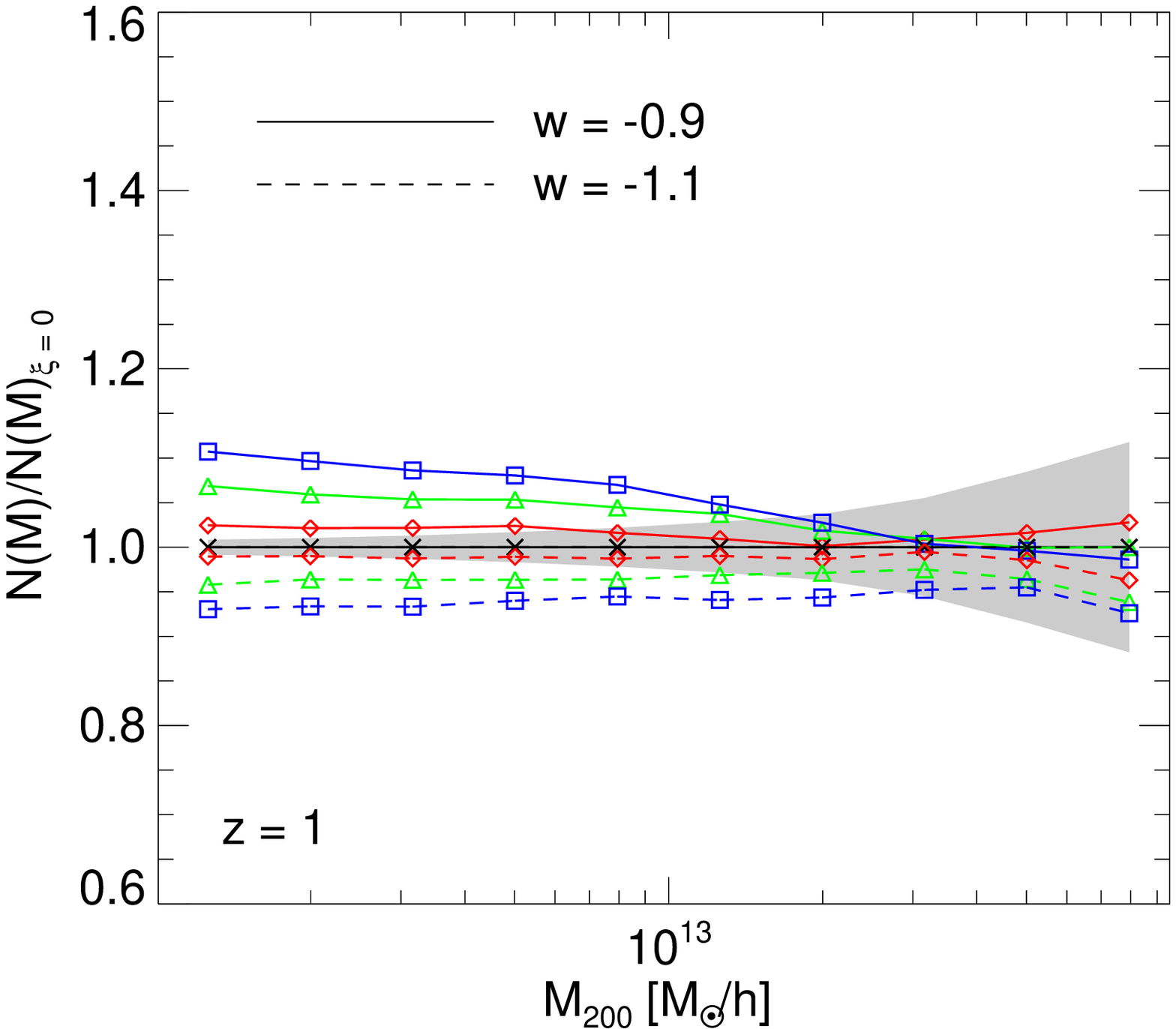}
\caption{{ The mass function ratio to the $\xi = 0$ case for all the models with $w_{\rm DE} \neq -1$ at three different redshifts $z=0$ ({\em left panel}), $z=0.5$ ({\em middle panel}), and $z=1$ ({\em right panel}). Solid lines refer to the $w_{\rm DE} = -0.9$ case while dashed lines refer to the $w_{\rm DE} = -1.1$ models, while the different colours and open symbols refer to the different values of the parameter $\xi $. The effect of enhancement or suppression of the halo abundance is almost mass-independent for all the models at $z=0$, while at higher redshifts the large masses appear less affected than the low masses.}}
\label{fig:HMF_ratio}
\end{figure*}

\begin{figure*}
\includegraphics[scale=0.3]{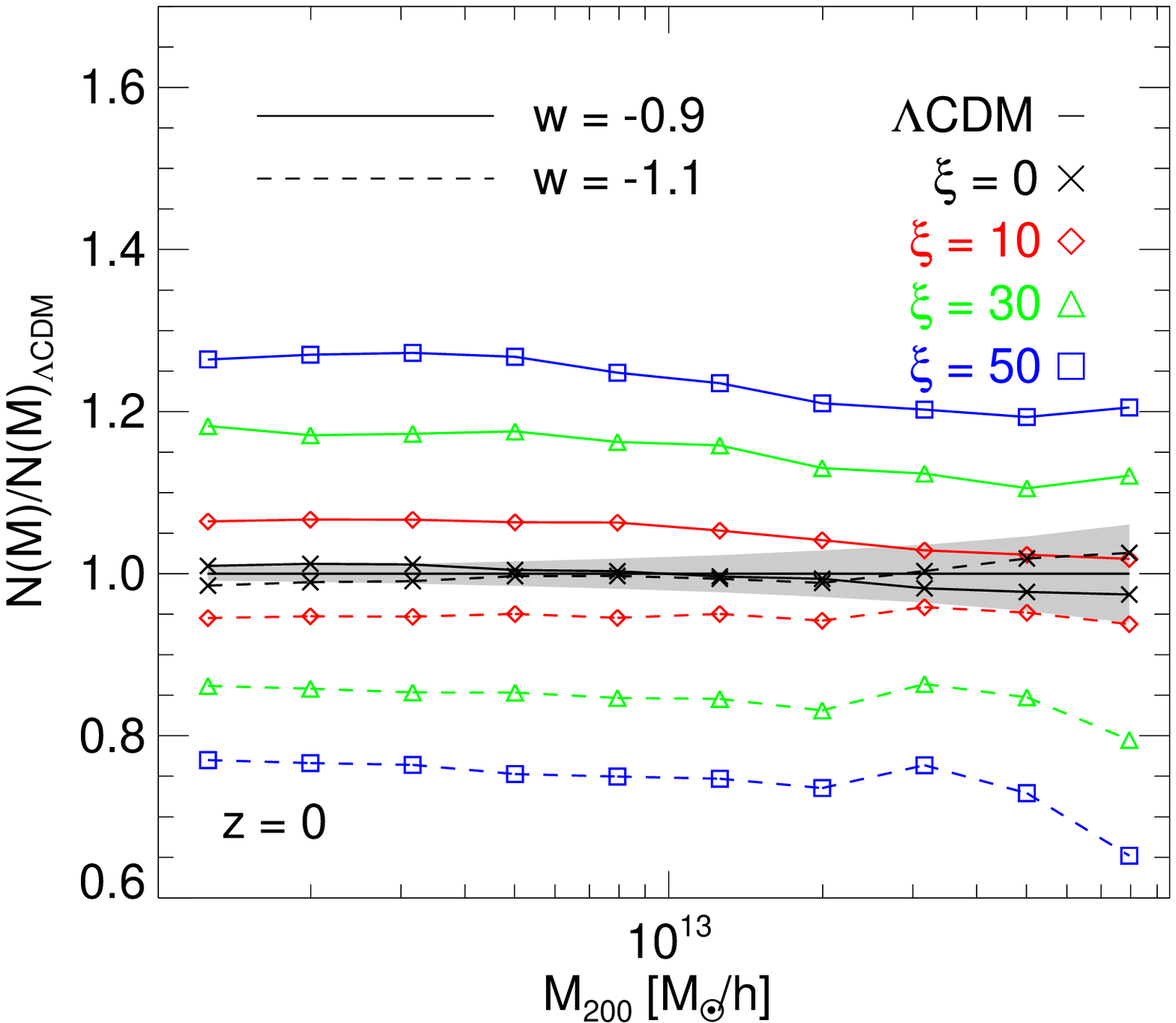}
\includegraphics[scale=0.3]{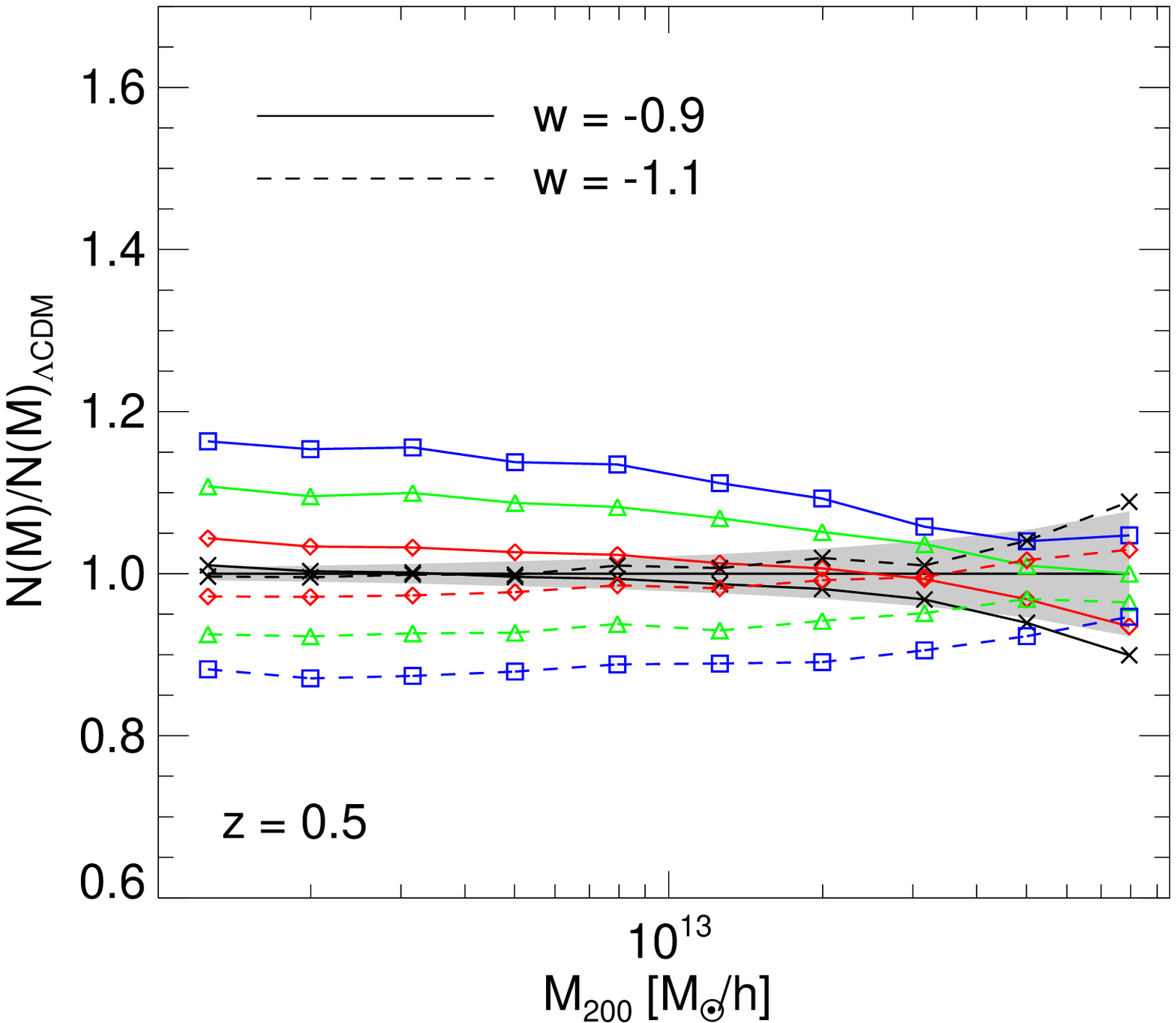}
\includegraphics[scale=0.3]{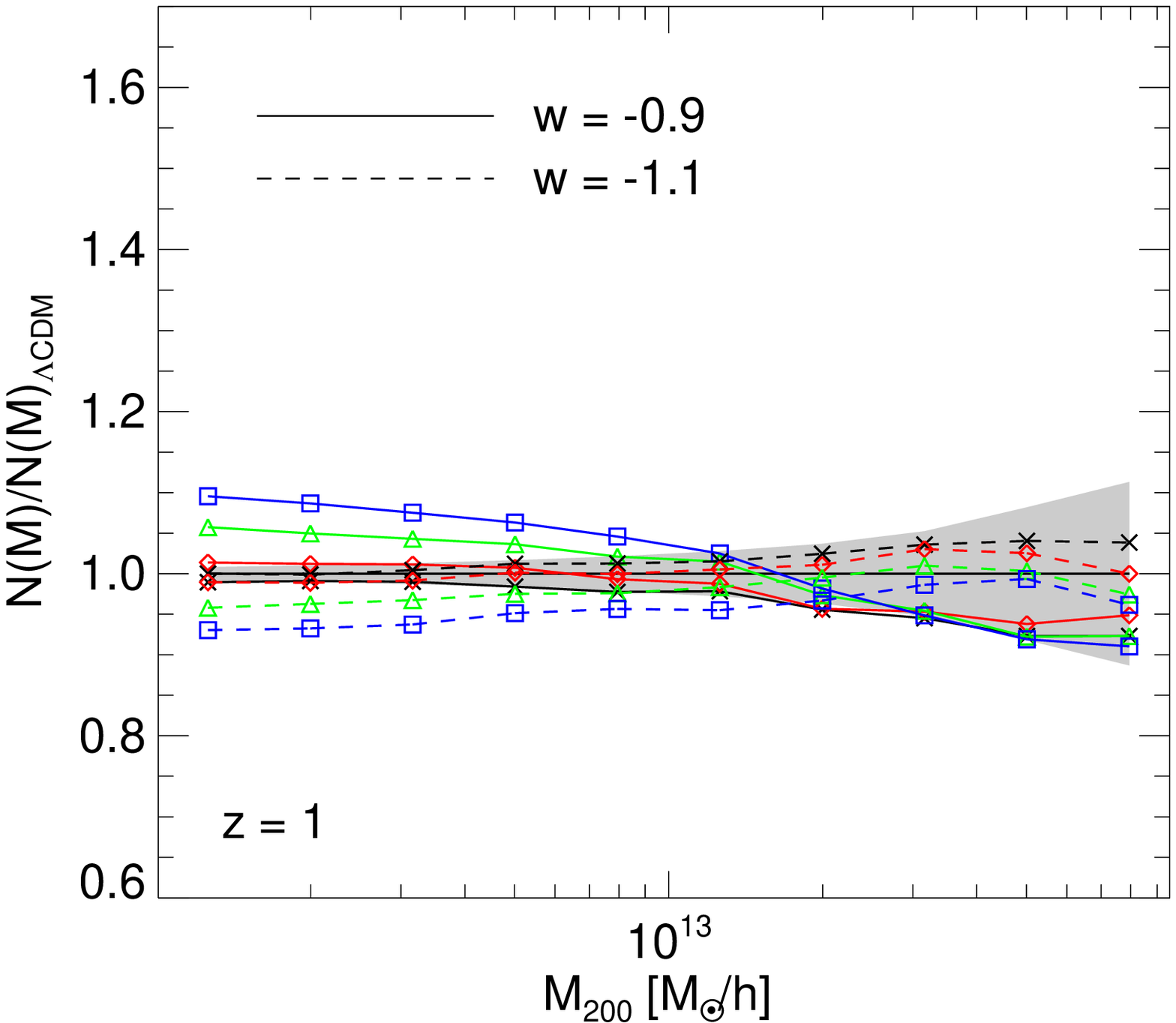}
\caption{{ The mass function ratio to the $\Lambda $CDM case for all the models under investigation in the present work. The three panels, colours, symbols, and line styles are the same as displayed in Fig.~\ref{fig:HMF_ratio}. Also in this case the effect of enhancement or suppression appears almost scale-independent for all the models at $z=0$ while for higher redshift the mass dependence is more pronounced than in the comparison among models with the same expansion history. At large masses at high redshifts it is also possible to observe a suppression of the abundance of large-mass halos for $w=-0.9$ models which did not occur at fixed expansion history.}}
\label{fig:HMF_ratio_LCDM}
\end{figure*}

With our halo catalogs at hand we have computed -- for each cosmology -- the halo mass function as the number of halos with virial mass $M_{200}$ lying within a series of $10$ logarithmically equispaced mass bins in the range $10^{12}-10^{14}$ M$_{\odot }/h$. This allows us to have a statistically significant sample of halos for each mass bin, with a minimum number of $75$ halos for the poorest bin used in our analysis (corresponding to the most massive bin in the $w=-1.1\,, \xi = 50\, [{\rm bn} \cdot {\rm c}^{3}/{\rm GeV}]$ model).

Similarly to what was shown above for the nonlinear matter power spectrum, in Figs.~\ref{fig:HMF_ratio} and \ref{fig:HMF_ratio_LCDM} we display the ratio of the halo mass function of all the models under investigation to the $\xi = 0$ case (for fixed expansion history), and to the standard $\Lambda $CDM cosmology, respectively. In both figures the grey shaded regions represent the Poissonian error on the ratio based on the number of objects included in each bin of the reference simulation, which for Fig.~\ref{fig:HMF_ratio} is taken to be the $w_{\rm DE} = -0.9\,, \xi = 50\, [{\rm bn} \cdot {\rm c}^{3}/{\rm GeV}]$ model. Also in this case, the two comparisons are qualitatively similar, although some differences between the two appear at the largest masses as a consequence of the exponential dependence of the halo mass function on the linear perturbations amplitude. This changes from model to model as a consequence of the different expansion history associated with the three possible different values of $w_{\rm DE}$. However, in both cases we observe a significant enhancement (suppression) of the abundance of halos at all masses within our mass range for the $w_{\rm DE} > -1$ ($w_{\rm DE}<-1$) models, with the magnitude of the effect increasing with the value of $\xi $, and with decreasing redshift. 

This turns out to be another highly distinctive feature of the DE-CDM scattering. In fact, while most non-standard cosmologies, including e.g. DE \citep[see e.g.][]{Courtin_etal_2011,Cui_Baldi_Borgani_2012} and Modified Gravity \citep[see e.g.][]{Baldi_etal_2013,Lombriser_etal_2013} scenarios, as well as primordial non-Gaussianity \citep[see e.g.][]{Grossi_etal_2007,Wagner_Verde_Boubekeur_2010}, massive neutrinos \citep[see e.g.][]{Castorina_etal_2013} and Warm Dark Matter \citep[see e.g.][]{Angulo_Hahn_Abel_2013} models, all affect the halo mass function with a specific mass dependence -- in most cases having a stronger impact on the high-mass tail -- the effect of the DE-CDM scattering appears to have a very weak dependence on the halo mass resulting in a roughly constant enhancement (or suppression) of the halo abundance over a wide mass range, at least at low redshifts. Some more pronounced mass dependence appears at higher redshifts where the most massive halos are less affected by the scattering as compared to the low mass ones. The maximum relative deviation from the $\Lambda $CDM model is obtained for the two $\xi = 50\, [{\rm bn} \cdot {\rm c}^{3}/{\rm GeV}]$ models at $z=0$ and ranges between 20 and 25\%. 

This is the result of the superposition of two distinct effects. On one side, the weak  modulation of the linear perturbations amplitude is expected to slightly suppress (enhance) the abundance of halos at large masses for $w_{\rm DE} > -1$ ($w_{\rm DE} < -1$) as a consequence of the exponential dependence of the high-mass tail of the halo mass function on $\sigma _{8}$. On the other side, the strong distortion of the matter power spectrum at nonlinear scales is expected to increase (reduce) the abundance of objects of all masses for $w_{\rm DE} > -1$ ($w_{\rm DE} < -1$) by changing the value of $M_{200}$ as a consequence of the dynamical change of the halo concentrations, as discussed above. Remarkably, the overall effect appears to be almost mass-independent, at least at low redshifts, while at higher redshifts the low-mass end of the halo mass function is more significantly affected.

\subsection{Halo velocity dispersion}

\begin{figure*}
\includegraphics[scale=0.3]{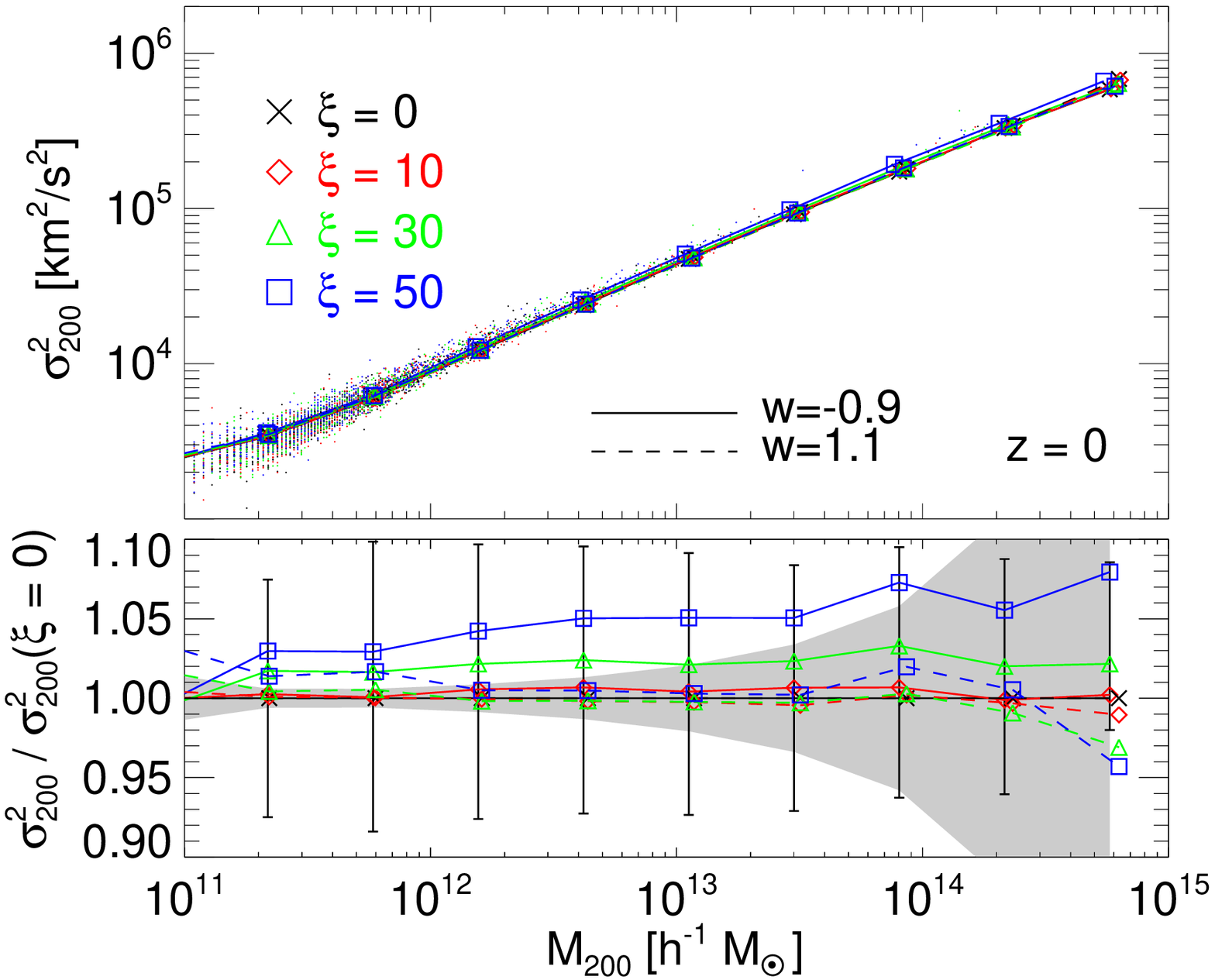}
\includegraphics[scale=0.3]{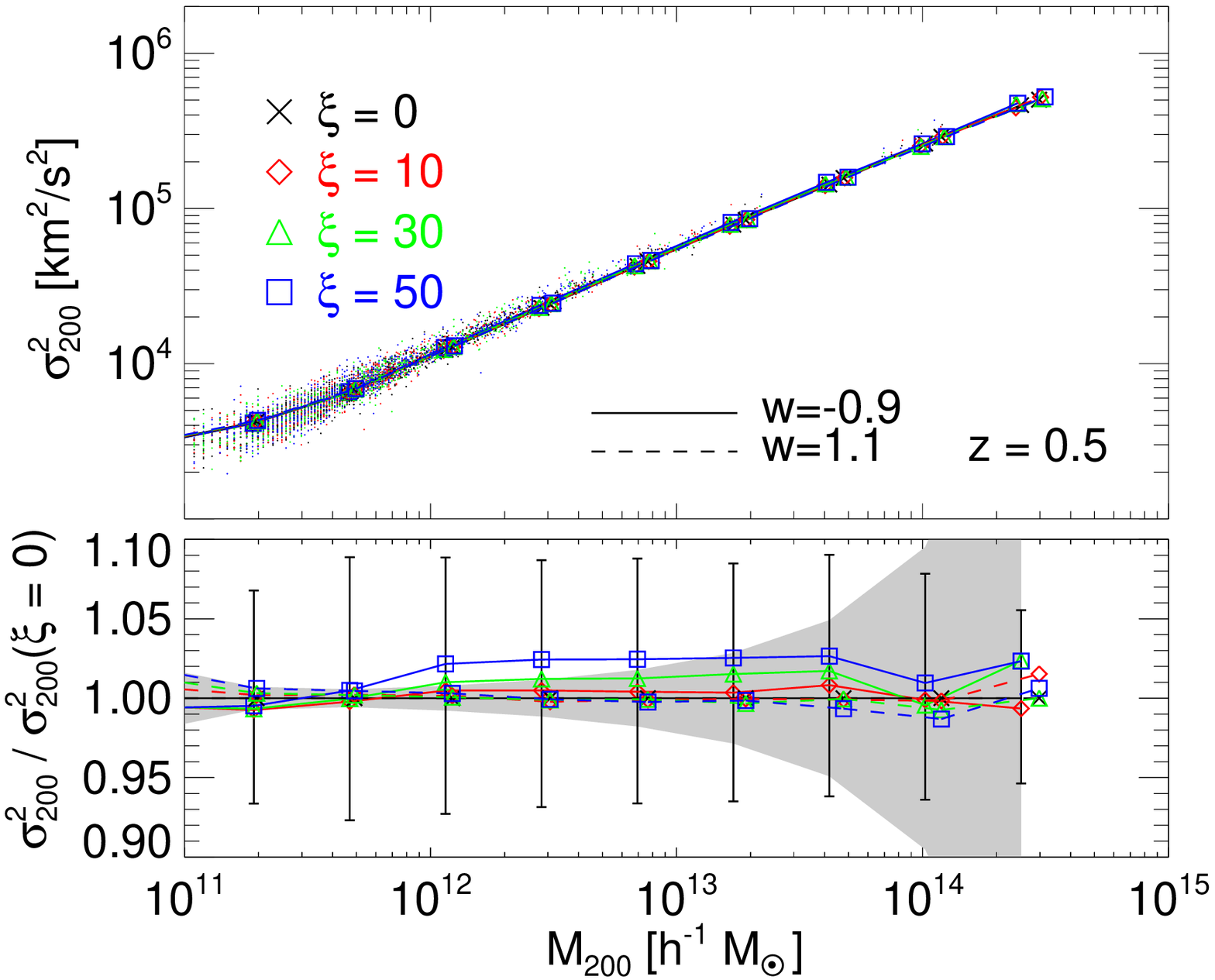}
\includegraphics[scale=0.3]{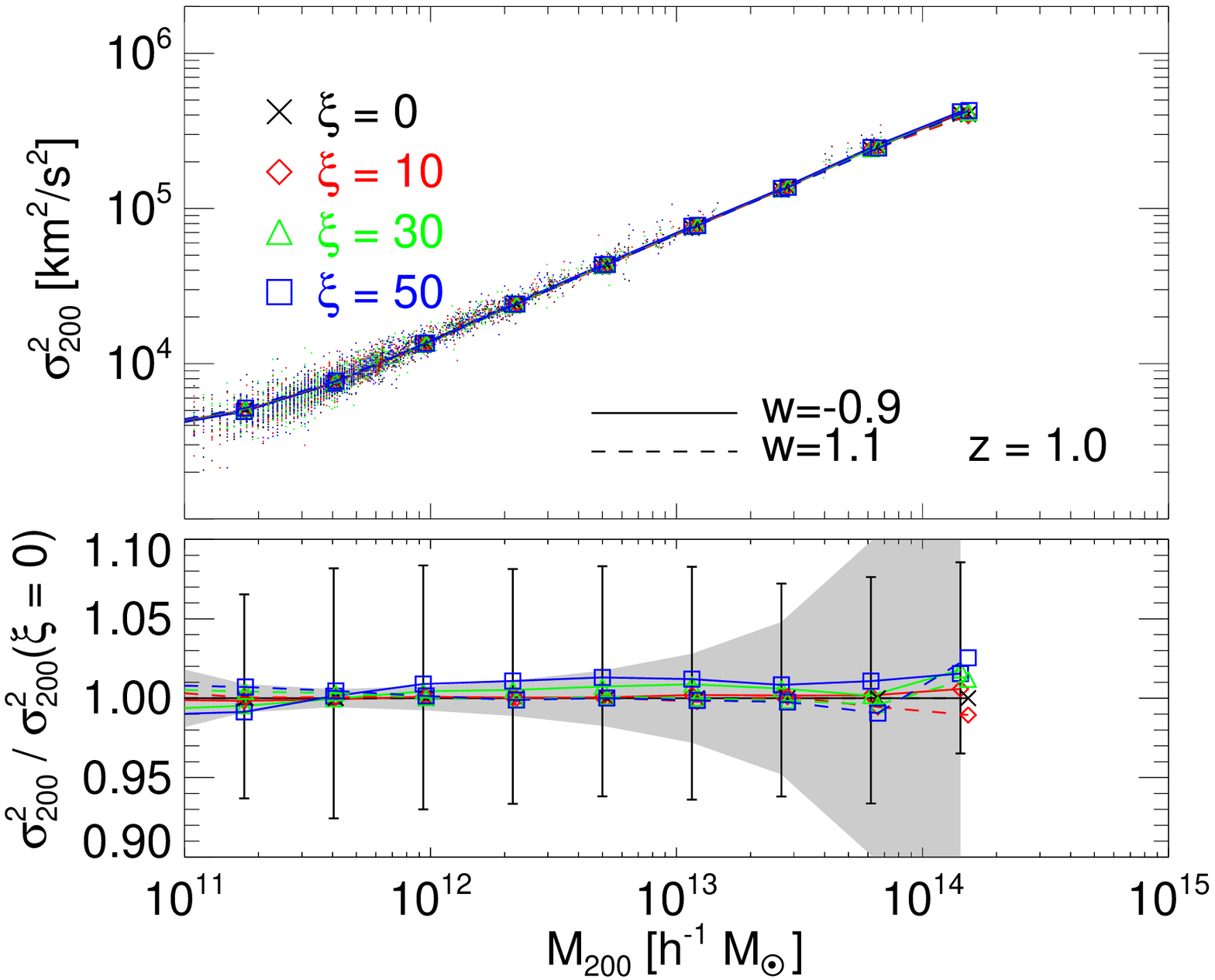}
\caption{{The one-dimensional velocity dispersion $\sigma^{2}_{200}$ as a function of halo mass $M_{200}$ and its ratio to the $\xi = 0$ model at the usual three different redshifts $z=0$ ({\em left}), $z=0.5$ ({\em middle}), and $z=1$ ({\em right}). In the upper panels we display as coloured dots a random subsample of the halos of each model and as solid and dashed lines the binned mean of the data points for the $w_{\rm DE} = -0.9$ and $w_{\rm DE} = -1.1$, respectively. In the lower panels we show the binned ratio to the non scattering model with identical expansion history.}}
\label{fig:velocity_dispersion}
\end{figure*}

As a further statistics of the cosmic structures properties in the presence of a DE-CDM scattering we investigate the relation between the one-dimensional velocity dispersion $\sigma ^{2}$ and $M_{200}$ for all the halos in our catalogs, and compare the results to the case of no scattering $\xi = 0$. This comparison is shown in Fig.~\ref{fig:velocity_dispersion} for models with fixed expansion history at the usual three different redshifts $z=0$, $z=0.5$, and $z=1$. In the upper panels the coloured dots represent a random subsample of all the halos in the catalogs while the solid  and dashed lines trace the mean value of $\sigma ^{2}$ within 10 logarithmically equispaced mass bins for the $w_{\rm DE}=-0.9$ and the $w_{\rm DE}=-1.1$ models, respectively. In the bottom panels we display the ratio of the binned average 1-D velocity dispersion of each model to the non scattering case with identical expansion history, again shown as solid (dashed) lines for $w_{\rm DE}=-0.9$ ($w_{\rm DE}=-1.1$). 

Also for these figures, the grey shaded region indicates the Poissonian error associated with the abundance of halos in the different bins of the reference simulation. As one can see from the plots, the models with $w_{\rm DE}=-0.9$ show a systematic enhancement of the 1-D velocity dispersion with respect to the $\xi = 0$ case over the whole mass range covered by our sample, with a weak mass dependence giving rise to a slightly stronger effect for the largest masses probed by our catalogs. The enhancement increases for increasing values of $\xi $ and for later times, reaching a maximum value of $\approx 7-8\%$ for the most massive halos in the $\xi = 50$ [bn $\cdot c^{3}/$GeV] model at $z=0$. This effect can be again interpreted as a consequence of the DE-CDM scattering on the structural properties of the halos: when a drag term is acting on the dark matter particles,  the halos are more concentrated and therefore our definition of $R_{200}$ moves inwards. Therefore, for a fixed halo mass, the potential well at  $R_{200}$ is deeper, which corresponds to an enhanced velocity dispersion.

Interestingly, we do not  find a similar effect in the opposite direction for the models with $w_{\rm DE}=-1.1$, as it was always the case for all the other observables investigated in this work. On the contrary, all the $w_{\rm DE}=-1.1$ cosmologies show very little deviations from their reference model, never exceeding $\approx 3-4\%$ even for the largest value of the $\xi $ parameter. 
Such different efficiency of the scattering in changing the halo velocity dispersion for {\em quintessence} and {\em phantom} expansion histories might be related to the later onset of the extra-scattering term occurring for the latter models, which  is clearly visible in Fig.~\ref{fig:drag_force}.
A detailed investigation of this effect would require a larger statistical sample and a higher resolution than is allowed by our present simulations, and is left for future work. It is nonetheless interesting to notice here that the absence of any impact of the DE-CDM scattering on the halo velocity dispersion for $w_{\rm DE} < -1$, if confirmed by future investigations, might provide a way to break possible degeneracies present in other observables between the signature of the scattering and the effects of the background expansion history.

\subsection{Halo bias}

\begin{figure}
\includegraphics[width=\columnwidth]{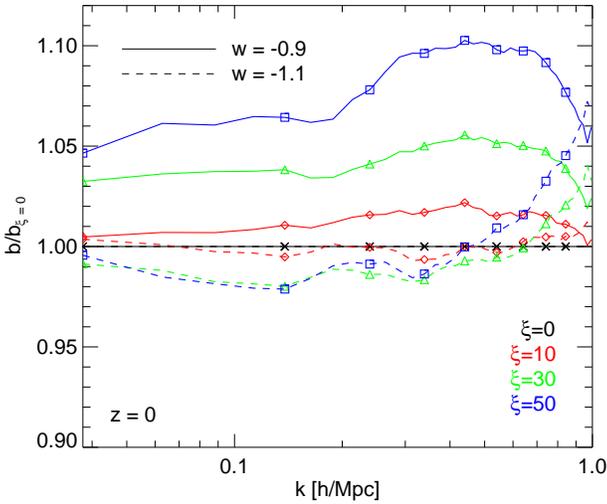}
\caption{The ratio of the halo bias to the non-scattering reference model with identical expansion history for both $w_{\rm DE} = -0.9$ ({\em solid lines}) and $w=-1.1$ ({\em dashed lines}).}
\label{fig:bias_ratio}
\end{figure}

We conclude our analysis by investigating the halo bias in our set of dark scattering cosmologies. We compute the bias by taking the ratio of the halo-matter cross-power spectrum $P_{\rm hm}(k)$  to the matter-matter power spectrum $P_{\rm mm}(k)$ for halos with mass above $3\times 10^{12}$ M$_{\odot }/h$. In Fig.~\ref{fig:bias_ratio} we show the ratio of the halo bias of all the scattering DE models to the corresponding non-scattering reference model with identical expansion history in the range of scales $0.04\leq k  \cdot h/{\rm Mpc} \leq 1$. Interestingly, also in this case the $w_{\rm DE}=-0.9$ models show a significant deviation with respect to the non-scattering scenario, with an increase of the halo bias a all scales, while the $w_{\rm DE}=-1.1$ models have much smaller deviations. The increase of the bias for the $w_{\rm DE} = -0.9$ models also shows a clear enhancement at the crossing between linear and nonlinear scales, reflecting the sharp transition displayed in Figure~\ref{fig:power_ratio} for the matter power spectrum. Also in this case, we argue that the different behaviour of the $w_{\rm DE} > -1$ and $w_{\rm DE} < -1$ models might be due to the later onset of the extra-scattering term for a {\em phantom} expansion history, as shown in Fig.~\ref{fig:drag_force}. We leave for future work a more detailed investigation of how the redshift evolution of the extra-force (which is in turn dictated by the evolution of the DE density) is related to the impact of the scattering on both the halo velocity dispersion and the bias. In particular, for models of EDE where the DE density never vanishes at high redshifts we would expect -- for a fixed scattering parameter $\xi $ -- a direct relation between the amount of EDE and the efficiency of the scattering in altering both the halo velocity dispersion and the bias.

\section{Conclusions}
\label{sec:concl}

In the present work we have investigated for the first time the observational effects of an elastic scattering between Cold Dark Matter particles and a perfect fluid associated with the cosmic Dark Energy in the nonlinear regime of structure formation by means of a suite of intermediate-resolution N-body simulations. As a first numerical investigation of these scenarios, we have restricted our analysis to the simplified case of a constant Dark Energy equation of state parameter taking both a canonical value $w_{\rm DE} = -0.9$ and a phantom-like value $w_{\rm DE} = -1.1$ and for each of these cases we have run four simulations with different values of the dimensional parameter $\xi $ associated with the ratio between the Dark Energy-Cold Dark Matter scattering cross section $\sigma _{c}$ and the CDM particle mass $m_{\rm CDM}$, namely $\xi = \left\{ 0\,, 10\,, 30\,, 50\right\}$ [bn $\cdot c^{3}/$GeV]. A further simulation for the standard $\Lambda $CDM cosmology has been run as reference. All simulations shared the same initial conditions, thereby discarding possible effects of the scattering before the starting redshift of the runs $z_{i}=99$, which we verified to be a safe approximation for the specific models under investigation. 

We analysed our simulations suite through a series of basic statistical properties of the large-scale matter distribution and structural properties of collapsed halos. More specifically, the main results of our work can be summarised as follows:
\begin{itemize}
\item {\bf Matter Power Spectrum} -- By comparing the full nonlinear matter power spectrum extracted from our simulations both to the standard $\Lambda $CDM reference and to the non-scattering scenario with identical expansion history (i.e. to the $\xi = 0$ run with the same $w_{\rm DE}$ for each model) we have shown that the power spectrum is  affected by the scattering process in a radically different way between the linear and the nonlinear regime of structure formation. In the former, a scale-independent shift in the power amplitude appears at linear scales, resulting in a suppression or an enhancement of the linear perturbations amplitude for $w_{\rm DE} > -1$ or $w_{\rm DE} > -1$, respectively, with a maximum relative difference of $\approx 12\%$ for the most extreme scenario. In the latter, instead, a strongly scale-dependent deviation with a much larger amplitude and an opposite sign as compared to the linear regime is found for scales below $k \sim 0.7-3\, h/$Mpc depending on the model and on the redshift. Such behaviour can be interpreted as a consequence of the injection or dissipation of kinetic energy within virialized (or virializing) structures, thereby altering the efficiency of the gravitational collapse.\\
\item {\bf Halo Concentrations} -- We computed the concentration of all the main substructures of our halo catalogs and compared the concentration-mass relation of all the models over the mass range $\approx 10^{12}-10^{15}$ M$_{\odot }/h$. We then compared this relation to the corresponding one for the case of no scattering ($\xi = 0$) and identical expansion history. As a general trend, we found that the scattering between Cold Dark Matter particles and the Dark Energy fluid induces an increase of the normalisation of the $c-M$ relation for $w_{\rm DE} > -1$ and correspondingly a decrease of the normalisation for $w_{\rm DE} < -1$. The effect is largest at $z=0$ and rapidly decreases with redshift, with a maximum deviation of a factor $\approx 2$ for the most extreme models ($\xi = 50$ bn $\cdot c^{3}/$GeV) at $z=0$. Such evolution can also be ascribed to the nonlinear effect of the extra-drag term associated with the elastic scattering, which determines an injection or a dissipation of kinetic energy from bound structures for $w < -1$ and $w > -1$, respectively, with a consequent expansion or contraction of the halos resulting in a lower or higher concentration for a fixed halo mass.\\
\item {\bf Halo Mass Function} -- For all our simulations we compared the abundance of halos in the mass range $10^{12} - 10^{14}$ M$_{\odot }/h$ to both the non scattering model with the identical expansion history and the standard $\Lambda $CDM cosmology. In both cases the scattering results in a significant increase of the halo abundance over the whole mass range for $w > -1$ and a corresponding decrease of the abundance for $w < -1$. Also in this case, the effect is largest at $z=0$ and rapidly decreases with redshift, with a maximum relative deviation of about $20\%$ at $z=0$ for the most extreme realisation of elastic scattering. In the comparison with the $\Lambda $CDM cosmology, we observe the superposition of the effects due to the scattering and to the different expansion history, especially at higher redshifts when the impact of the scattering term is significantly weaker than at later epochs, and the high-mass tail of the halo mass function shows differences from model to model that are mainly related to the exponential dependence of the multiplicity function on the underlying value of the $\sigma _{8}$ normalisation.\\
\item {\bf Velocity dispersions} --  We investigated the relation between the one-dimensional velocity dispersion of the halos and their mass in the various models under study, and compared the obtained relation to the case of no scattering with identical cosmological expansion history. Our results indicate a systematic increase of the velocity dispersion at fixed mass for increasing values of the scattering parameter $\xi $ in the $w_{\rm DE} > -1$ scenarios. The effect is again largest at low redshifts and shows a slightly positive correlation with the halo mass, reaching a maximum value of $\approx 8\%$ for the largest masses in the most extreme model. Interestingly, the same feature does not appear for the $w < -1$ models which show no significant deviations from the non-scattering case even for large values of the scattering parameter $\xi $.  
\\
\item {\bf Halo Bias} -- We compared the halo bias of all the dark scattering models under investigation to the corresponding non-scattering model with identical expansion history, finding a systematic increase of the bias at all scales for the $w_{\rm DE} = -0.9$ cosmologies, for increasing values of the scattering parameter $\xi $, with a clear additional enhancement occurring at the transition between linear and nonlinear scales reflecting the transition observed in the matter power spectrum. Also in this case, the $w_{\rm DE} = -1.1$ scenarios show much smaller deviations, similarly to what we found for the halo velocity dispersion.
This starkly different behaviour of the ``canonical" and ``phantom" Dark Energy scenarios in the presence of a scattering with the Cold Dark Matter particles deserves further investigations, and will be addressed in future works.
\end{itemize}
\ \\

To conclude, we have presented the first investigation of nonlinear structure formation in the context of cosmological models featuring a non-vanishing scattering cross section $\sigma _{c}$ between Cold Dark Matter particles and a Dark Energy fluid characterised by a constant equation of state parameter with either a ``canonical" or a ``phantom" behaviour. Our results, based on the first N-body simulations of this class of models ever performed, indicate that the nonlinear evolution of cosmic structures might provide significantly tighter constraints on a possible process of elastic scattering in the dark sector as compared to the present bounds based on linear observables only. A more realistic investigation of the effects of the scattering in the presence of a time-dependent Dark Energy equation of state, as for e.g. Early Dark Energy cosmologies, is presently ongoing and will be discussed in an upcoming paper.

\section*{Acknowledgments}
We are deeply thankful to Licia Verde and Raul Jimenez for their helpful comments on the draft.
MB is supported by the  Marie Curie Intra European Fellowship
``SIDUN"  within the 7th Framework Programme of the European Commission. FS acknowledges support from the European Research Council under the Seventh Framework Programme FP7-IDEAS-Phys.LSS 240117.
The numerical simulations presented in this work have been performed 
and analysed on the Hydra cluster at the RZG supercomputing centre in Garching.

\bibliographystyle{mnras}
\bibliography{dm_scatter.bib,baldi_bibliography.bib}
\label{lastpage}

\end{document}